\documentclass[journal=jacsat,manuscript=article]{achemso}

\usepackage[version=3]{mhchem} 
\setkeys{acs}{keywords = true} %to show keywords
\usepackage{graphicx}% Include figure files
\graphicspath{{images/}}
\SectionNumbersOn

\usepackage{subcaption}
\usepackage{dcolumn}% Align table columns on decimal point
\usepackage{tabularx,booktabs}
\usepackage{bm}
\usepackage{xcolor}
\usepackage{color, colortbl}
\usepackage{ulem}
\definecolor{Gray}{gray}{0.9}

\title{Modeling Chemical Exfoliation of Non-van der Waals Chromium Sulfides by Machine Learning Interatomic Potentials and Monte Carlo Simulations}
\author{Akram Ibrahim}
\affiliation{%
 Department of Physics, University of Maryland Baltimore County, 1000 Hilltop Circle, Baltimore, MD 21250, USA
}%
\author{Daniel Wines}
 \affiliation{%
 Material Measurement Laboratory, National Institute of Standards and Technology,
Gaithersburg, MD 20899, USA 
}%
\alsoaffiliation{%
 Department of Physics, University of Maryland Baltimore County, 1000 Hilltop Circle, Baltimore, MD 21250, USA
}%
 
\author{Can Ataca}
\affiliation{%
 Department of Physics, University of Maryland Baltimore County, 1000 Hilltop Circle, Baltimore, MD 21250, USA
}%

\email{ataca@umbc.edu}
\date{\today}


\begin{document}

\begin{abstract}

The chemical exfoliation of non-van der Waals (vdW) materials to ultrathin nanosheets remains a formidable challenge. This difficulty arises from the strong preference of these materials to engage in three-dimensional chemical bonding, resulting in uncontrolled atomic migration into the vdW gaps during cation deintercalation from the bulk structure, ultimately leading to unpredictable structural disorder. Computational models capable of comprehending the widespread nonstoichiometric local environments resulting from disordered atomic migrations during exfoliation of non-vdW materials are crucial for understanding the underlying mechanisms. Here, we propose a generic framework using neural network potentials (NNPs) to accurately model nonstoichiometric systems over a broad range of vacancy concentrations. We apply our framework to investigate the crystal structures and phase transformations occurring during the exfoliation of non-vdW nonstoichiometric Cr$_{(1-x)}$S systems, a compelling material category with substantial potential for two-dimensional (2D) magnetic applications. The efficacy of the NNP outperforms the conventional cluster expansion, exhibiting superior accuracy and transferability to unexplored crystal structures and compositions. By employing the NNP in simulated annealing optimizations, we predict low-energy Cr$_{(1-x)}$S structures anticipated to result from experimental synthesis. A notable structural transition is discerned at the Cr$_{0.5}$S composition, with half of the Cr atoms preferentially migrating to vdW gaps. This aligns with experimental observations in the chemical exfoliation of 2D CrS$_2$, and emphasizes the vital role of excess Cr atoms beyond the Cr/S = $1/2$ composition ratio in stabilizing vdW gaps. Additionally, we utilize the NNP in a large-scale vacancy diffusion Monte Carlo simulation to illustrate the impact of lateral compressive strains in catalyzing the formation of vdW gaps within non-vdW CrS$_2$ slabs through Poisson’s axial expansion. This provides a direct pathway for more facile exfoliation of ultrathin nanosheets from non-vdW materials through strain engineering. The implemented methodology, leveraging machine learning potentials, is imperative to bridge the quantum-level accuracy to large scales necessary for modeling the intricate mechanisms underlying the chemical exfoliation of non-vdW materials.

\end{abstract}

\maketitle

%\tableofcontents

\section{Introduction}

Recent years have witnessed significant progress in the synthesis of two-dimensional (2D) nanosheets derived from materials that lack a layered van der Waals (vdW) parent structure, broadening the scope of exfoliation research to encompass a wide variety of previously unexplored 2D non-vdW materials \cite{jin2021two, balan2022non}. Among the family of non-vdW materials, Cr-based compounds, particularly magnetic non-vdW Cr-S materials, have garnered substantial attention within the field of $2$D magnetic materials with a myriad of potential applications in electronic and spintronic devices \cite{hasan2023crystal, gibertini2019magnetic, jin2021two}, especially following the discovery of $2$D magnets like CrI$_3$ and CrGeTe$_3$ \cite{lado2017origin, yang2020solution}. The Cr-S system comprises a minimum of six distinct known phases, namely CrS, Cr$_7$S$_8$, Cr$_5$S$_6$, Cr$_3$S$_4$, Cr$_2$S$_3$, and Cr$_5$S$_8$ \cite{song2019soft}, all exhibiting a hexagonal NiAs-type structure, with vacancies in the Cr sublattice responsible for the formation of the non-stoichiometric phases. Cr$_2$S$_3$ is the most stable phase, attributed to Cr's preference for the $3+$ oxidation state. Recently, nano-thick 2D Cr-S slabs have been successfully fabricated. For example, Cr$_2$S$_3$ flakes with a thickness ranging from tens of nanometers down to a single unit cell ($\sim 1.8$ nm) were synthesized using chemical vapor deposition (CVD) \cite{cui2020controlled, chu2019sub}. The resulting 2D Cr$_2$S$_3$ material is a semiconductor with ferrimagnetic ordering and a Néel temperature of around $120$ K. Additionally, a few-nanometer-thick 2D flakes of $1$T-CrS$_2$ were synthesized using CVD, exhibiting metallic behavior and room temperature ferromagnetism \cite{xiao2022van}. Moreover, a semiconducting H$_x$CrS$_2$ crystalline/amorphous layered material, exhibiting antiferromagnetic ordering and a Néel temperature of $5$ K, was synthesized with a thickness of $2$-$3$ nm using chemical exfoliation \cite{song2019soft}.

The strong interlayer binding inherent in bulk non-vdW layered materials constitute a formidable barrier to conventional mechanical exfoliation methods. Consequently, alternative approaches such as chemical exfoliation or direct liquid-phase exfoliation (LPE) are commonly employed for the top-down synthesis of non-vdW materials. Chemical exfoliation is especially pertinent for crystal structures that do not naturally exist in pristine bulk form, whether as vdW or non-vdW structures \cite{song2021properties}. This is particularly evident in the case of Cr dichalcogenides, i.e., CrX$_2$ (where X = S, Se, and Te), which do not naturally exist in a bulk transition metal dichalcogenide (TMD) layered or pyrite form \cite{song2019soft}. Nevertheless, these structures can be identified in other ternary compounds, such as ACrX$_2$ (where A = Li, Na, and K) \cite{ushakov2013magnetism, song2019soft, song2021kinetics}. These ternary compounds can serve as base structures for synthesizing 2D Cr-based TMDs through chemical exfoliation. The initial step in chemical exfoliation involves the removal of extraneous cations present in the base bulk structure, either through cation exchange or oxidative removal. For instance, proton exchange with Na in NaCrS$_2$ using HCl/ethanol solution \cite{song2019soft}, K deintercalation from both KCrSe$_2$ \cite{van1980crse2, song2021kinetics} and KCrTe$_2$ \cite{freitas2015ferromagnetism} through I$_2$ oxidation, have been employed during the chemical exfoliation of CrS$_2$, CrSe$_2$, and CrTe$_2$ structures, respectively. The second step of chemical exfoliation involves LPE to induce fragmentation and delamination of atomic layers from the formed bulk structures. However, owing to the strong tendency of non-vdW materials to form three-dimensional (3D) chemical bonds, they undergo uncontrolled cation migration from the layers (Cr atoms within CrS$_2$ layers in this context) to the vdW gaps during the initial deintercalation process, thereby retaining their non-vdW characteristics \cite{song2019soft, stiles2023unlocking, song2021properties, song2021kinetics}. As a consequence, the exfoliated $2$D sheets typically exhibit thicknesses on the order of several nanometers, thereby impeding the realization of the distinct properties observable in monolayer-thin sheets, characterized by enhanced quantum confinement and reduced interlayer interactions. As a result, there is a growing interest in developing computational methods to achieve a deeper understanding and, consequently, greater control over the chemical exfoliation of ultrathin nanosheets from these non-vdW crystals.

Nonstoichiometric local atomic environments, characterized by varying elevated vacancy concentrations, are commonly observed in chemical exfoliation processes \cite{song2019soft, song2021properties}. The ubiquity of these nonstoichiometric environments can be attributed to the disordered atomic migration that occurs following the deintercalation process associated with the chemical exfoliation of non-vdW materials \cite{song2019soft}. This phenomenon leads to distinct compositional variations that are evident both temporally and spatially. Consequently, to gain insights into the intricate mechanisms underlying the chemical exfoliation of non-vdW materials, it is imperative to develop accurate computational models capable of comprehending nonstoichiometric environments across a wide range of vacancy concentrations. Modeling nonstoichiometric systems exclusively through density functional theory (DFT) proves infeasible due to the extensive spectrum of conceivable compositions and atomic configurations, which escalates exponentially with the number of atoms in the unit cell \cite{oganov2011modern}. To mitigate this challenge, researchers commonly employ special quasi-random structure (SQS) cells \cite{zunger1990special, jiang2009first} as representative unit cells that effectively capture the statistical features of atomic disorder, enabling DFT predictions for compositions attainable within compact unit cells. To extrapolate to other compositions or facilitate large-scale simulations, the common practice is to employ a surrogate model like cluster expansion (CE) \cite{zhang2015self, ravi2010cluster, xie2023revealing}. While the on-lattice CE is well-suited for fixed-lattice alloy systems, nonstoichiometric systems frequently manifest significant deviations from the parent lattice, and exhibit notable sensitivity of atomic arrangements to mechanical strains \cite{haselmann2021negatively, aschauer2013strain, komsa2013point}. Therefore, a more dependable approach entails adopting an off-lattice model that incorporates both configurational and geometrical features of nonstoichiometric systems. Additionally, CE's transferability to unobserved on-lattice configurations and compositions remains a subject warranting further investigation \cite{xie2022machine}.

Machine learning potentials (MLPs) have recently emerged as highly promising tools in computational materials science due to their near-DFT accuracy, nearly linear scaling with system size, and exceptional transferability to diverse chemical environments. Prominent examples of MLPs include neural network potentials \cite{behler2007generalized, ko2021fourth, smith2017ani, zhang2018deep}, Gaussian approximation potentials \cite{bartok2010gaussian, bartok2013representing}, moment tensor potentials \cite{shapeev2016moment}, spectral neighbor analysis potentials \cite{thompson2015spectral}, atomic cluster expansion potential \cite{bochkarev2022efficient, lysogorskiy2021performant}, and graph neural network potentials \cite{batzner20223, chen2022universal, choudhary2023unified}. The high flexibility of MLPs allows for broad applicability across different types of matter, encompassing bulk \cite{bartok2018machine, shaidu2021systematic, mangold2020transferability} and 2D crystals \cite{mortazavi2020exploring}, amorphous materials \cite{deringer2018realistic, caro2018growth}, liquids \cite{morawietz2016van, li2021development, daru2022coupled}, interfaces \cite{artrith2019machine, achar2022using}, and clusters \cite{tang2023machine, jager2018machine}. In the domain of disordered systems, MLPs were employed to investigate binary alloys spanning a wide range of compositions \cite{rosenbrock2021machine, hodapp2021machine, nikoulis2021machine, andolina2021robust, li2022transferable}, high entropy alloys \cite{li2020complex, kostiuchenko2019impact, men2023understanding}, and grain boundaries \cite{wagih2020learning, lee2023atomic, yokoi2020neural, nishiyama2020application}. However, there exists a conspicuous gap in the literature concerning the application of MLPs to nonstoichiometric systems characterized by varying elevated vacancy concentrations. Herein, we address this gap by examining the efficacy of neural network potentials (NNPs) for modeling nonstoichiometric chromium sulfides, a material that has not been explored in the existing literature using either machine learning or conventional potentials. 

In this study, we present a general workflow for modeling nonstoichiometric systems using NNPs (Methods). The NNP's performance is demonstrated to surpass that of CE, showcasing near-DFT accuracy and remarkable transferability to unobserved structures exhibiting diverse compositions (Results – Subsection A). By incorporating the NNP into simulated annealing (SA) optimizations (Results – Subsection B), we gain insights into the low-energy ground-state and metastable Cr-S structures that are anticipated to result from experimental synthesis. The outcomes highlight the crucial role played by the surplus Cr atoms beyond the Cr/S $= 1/2$ ratio in stabilizing the vdW gaps.  Moreover, a notable structural transition is observed for the CrS$_2$ phase, characterized by a preferential migration of half of the Cr atoms to the vdW gaps. This finding aligns with experimental observations, illuminating the non-vdW nature of 2D CrS$_2$. Moreover, we leverage the NNP in a large-scale vacancy diffusion Monte Carlo (MC) simulation to emphasize the impact of lateral compressive strains in catalyzing the formation of vdW gaps within initially non-vdW CrS$_2$ slabs, paving a straightforward route for more facile exfoliation of ultrathin CrS$_2$ nanosheets through strain engineering (Results – Subsection C). The NNP's capacity to handle lattice strains, optimize atomic positions, and accurately consider spatially and temporally diverse nonstoichiometric compositions in large-scale simulations highlights the necessity of employing MLPs for modeling nonstoichiometric systems prevalent during chemical exfoliation of non-vdW materials. This task is unachievable solely with DFT and prone to considerable inaccuracies when attempted with conventional CE methods.

\section{Methods}

\begin{figure*}[btp] 
    \raggedright
    \includegraphics[width=\textwidth]{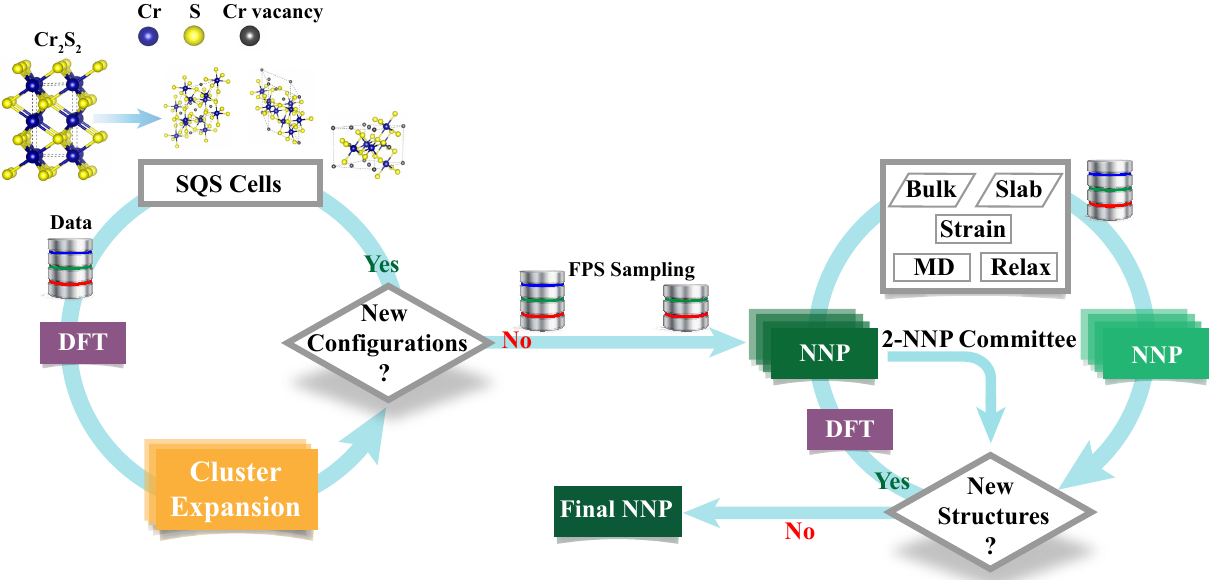}
    \caption{\textbf{The NNP generation workflow.} The first part (loop on the left) involves the utilization of the CE Hamiltonian to iteratively sample small SQS cells ($< 20$ atoms) toward ground states at $22$ distinct compositions. The NNP is first trained on snapshots, sampled using a FPS algorithm, from the DFT relaxations performed for CE-sampled SQS cells. In the second part (loop on the right), bulk structures and slab cuts for a subset of the identified ground states are subjected to strain around equilibrium volumes, and the NNP (dark green rectangle) is used to either run NVT-MD or fixed-volume relaxations. A comparison of energy and forces predictions is made with another variable-configuration NNP (light green rectangle) of slightly different hyperparameters (committee of $2$ NNP models) to sample extrapolative structures and iteratively update the dataset until convergence is achieved.} 
     \label{fig:main_figure}
\end{figure*}

\textbf{NNP model workflow.} Figure \ref{fig:main_figure} depicts the workflow employed for the construction of the NNP model, with comprehensive methodological details provided in the following subsections. The dataset generated through this workflow is summarized in Table S1. The NNP fitting procedure consists of two principal steps. Firstly, a CE Hamiltonian is iteratively fitted to SQS cells (referred to as N$_{SQS\;cells}$ in Table S1) generated for distinct vacancy concentrations within the Cr sublattice of the stoichiometric Cr$_2$S$_2$ hexagonal NiAs-type structure (space group P$6{_3}$/mmc). Thus, these structures can be represented by the formula unit Cr$_{(1-x)}$S, where $x$ denotes the Cr vacancy concentration. In this context, the Cr/vacancy sites are regarded as the active sites capable of accommodating 2 different species, while the S sites exclusively act as "spectators" within our model. The CE method, known for its simplicity and efficiency in sampling low-energy SQS cells that converge toward ground states, acts as an auxiliary model for exploring the structural space across various compositions before implementing the NNP. During CE fitting, DFT relaxations are conducted, and structural snapshots (denoted as $N_{SQS\;strucs}$ in Table S1) are sampled using a farthest point sampling (FPS) algorithm to ensure more uniform sampling along the relaxation path \cite{bartok2017machine, tirelli2022high}. These snapshots are then used to train a primary NNP model. This approach ensures that the NNP is trained on the equilibrium properties of the diverse atomic arrangements observed from the CE-sampled SQS cells, regardless of their relative energies with respect to the ground state.

Secondly, for an approximately uniform sample of these compositions, the identified ground state SQS cells are isotropically strained around the equilibrium volume. Subsequently, the NNP is utilized to perform volume-constrained energy minimization or conduct constant-temperature, constant-volume (NVT) molecular dynamics (MD) simulations. The NNP model is updated iteratively by selecting structural snapshots (referred to as $N_{GS\;bulk}$ in Table S1) from these calculations, using a committee of two NNP models \cite{schran2020committee, artrith2012high}. The configuration of the second NNP is changed over iterations to introduce dynamic variability in its predictions, thereby enhancing the identification of extrapolative structures. This process helps improve the predictability of elastic and vibrational properties for ground-state structures across different compositions. Additionally, for a few compositions near the ground-state composition (Cr$_{0.667}$S), $2$D slabs (denoted as $N_{GS\;slab}$ in Table S1) with a thickness of two unit cells and varying surface terminations are generated from the ground-state bulk SQS cells and added to the dataset. This augmentation enables the model to gain exposure to atomic environments of surface atoms across different surface compositions. Further details on the fitting of the CE and NNP models are presented in Table S2 and Table S3, respectively. To evaluate the predictive capability of the NNP, the root mean squared error (RMSE) is reported for both energies and forces, using $80/20$ train/test dataset split. For CE, we provide both the energy RMSE obtained through training on the entire SQS dataset, as well as the N-fold cross-validation (CV) score, where N denotes the number of distinct SQS cells in the dataset. After validation, both the final NNP and CE models are trained on their respective complete datasets to include all available data points.

\textbf{DFT calculations.} We use DFT for both training and validation of the CE and NNP models \cite{hohenberg1964inhomogeneous}. DFT calculations are performed within the Generalized Gradient Approximation (GGA) by Perdew, Burke, and Ernzerhof (PBE), using the Projector Augmented Wave (PAW) pseudopotentials implemented in the plane wave Vienna Ab initio Simulation Package (VASP) \cite{perdew1996generalized, blochl1994projector, kresse1996efficient}. Collinear spin-polarized calculations are adopted with plane waves basis cutoff at $350$ eV, and the integration on the first Brillouin zone is carried out on uniform Monkhorst-Pack meshes with a meshing density of $4000$ k-points per reciprocal atom \cite{monkhorst1976special}. The electronic self-consistent loop is ended if the energy change is less than $10^{-5}$ eV.

\textbf{NNP model fitting.} We employ the n2p2 neural network potential package \cite{singraber2019parallel, singraber2019library} for training our NNP model. The Large-scale Atomic/Molecular Massively Parallel Simulator (LAMMPS) package is used to perform local structural relaxations and MD simulations, using the fitted NNP as an input potential through the n2p2-LAMMPS interface \cite{thompson2022lammps, singraber2019library}. The ATAT package tools are utilized for fitting the CE model and enumeration of the SQS cells \cite{vanatat, van2019alloy, van2009multicomponent}.

We adopt the Behler-Parrinello high-dimensional NNP approach \cite{behler2007generalized}. This approach assumes that the total energy, $E$, of a structure can be represented as a sum of individual atomic energy contributions from constituent atoms. For our binary Cr-S system, this can be expressed as the sum of atomic energy contributions for the elements Cr and S. 
\begin{equation}
E=\sum_{i=1}^{N_{\text {Cr}}} E_{i} (\mathbf{G_i})+\sum_{i=1}^{N_{\text {S}}} E_{i}(\bf{G_i})
\end{equation}

The atomic energies, $E_i$, are generated by a set of feed-forward atomic neural networks (NNs). $E_i$ is a function of the local environment descriptor vector $\bf{G_i}$, which featurizes the atomic neighborhod within a given cutoff radius around atom $i$ and constitutes the input layer of the atomic NN. Specifically, our model comprises two distinct atomic NNs: one designed for Cr atoms and another for S atoms. The $\mathbf{G_i}$ vectors are constructed using the Behler-Parrinello symmetry functions (SFs) as descriptors \cite{behler2011atom}. The SFs consist of a radial function comprising sums of two-body Gaussians and an angular function incorporating additional sums of three-body terms.
\begin{equation}
G_{i}^{\text{rad}}=\sum_{j \neq i} \mathrm{e}^{-\eta\left(r_{i j}-r_{s}\right)^{2}} f_{c}\left(r_{i j}\right)
\label{eq:eq3}
\end{equation}

\begin{equation}
\begin{aligned}
G_{i}^{\text{ang}}=& 2^{1-\zeta} \sum_{\substack{j, k \neq i \\ j<k}}\left(1+\lambda \cos \theta_{i j k}\right)^{\zeta} \\
& \times \mathrm{e}^{-\eta\left(r_{i j}^{2}+r_{i k}^{2}\right)} f_{c}\left(r_{i j}\right) f_{c}\left(r_{i k}\right)
\end{aligned}
\end{equation}

Both SFs are many-body functions; they depend on the positions of all atoms within a cutoff radius, $r_c$. We use a cubic hyperbolic tangent cutoff function, $f_c$. The cutoff radii for radial and angular SFs, as well as the atomic NN architecture in our NNP model, are summarized in Table S3. Hyperbolic tangent is used as the activation function for hidden nodes, while the identity function serves as the activation for the output layer.

%\begin{equation}
%f_{c}\left(r_{i j}\right)=\left\{\begin{array}{cc}
%\tanh ^{3}\left(1-\frac{r}{r_{c}}\right) & r_{i j} \leq %r_{c} \\
%0 & r_{i j}>r_{c}
%\end{array}\right.
%\end{equation}

The hyperparameter $\lambda$ is set to $\{1, -1\}$, which shifts the maxima of the cosine function between $\theta_{ijk} = 0^\circ$ and $\theta_{ijk} = 180^\circ$ \cite{behler2011atom}. The $\zeta$ values are set to $\{1, 4, 24\}$ to cover multiple angular resolutions. Our SFs are centered at the central atom (i.e., $r_s=0$). The radial space around the central atom is smeared by $6$ different values of the hyperparameter $\eta$, which controls the radial resolution (Gaussian width). The $\eta$ values are determined using the Imbalzano method: $\eta_m = \left(\frac{n^{m/n}}{r_c}\right)^2$, where $n=5$ and $m=0, 1, .., n$ \cite{imbalzano2018automatic}.

The NNP weight optimization employs an adaptive Kalman filter, updating weight parameters with each energy or force component independently as individual pieces of information \cite{singraber2019parallel}. Forces are obtained as the analytical negative gradient of the energy function. Due to the abundance of force components compared to total energy entries, training is performed in each epoch on a pattern containing total energies of all structures and only a randomly-selected fraction of forces. The optimal composition corresponded to force updates approximately $2.6$ times the number of energy updates in each epoch.
%\begin{equation}
%\mathbf{F}_{i}=-\sum_{j} \sum_{k} \frac{\partial E_{j}}%{\partial G_{j k}} \frac{\partial G_{j k}}{\partial %\mathbf{R}_{i}}
%\end{equation}
%where the atom index $j$ runs over all the atoms within the distance $r_c$ from central atom $i$, and index $k$ runs over the descriptor components.
The NNP training procedure selects structures randomly but prioritizes more informative ones for weight updates. A structure is considered informative if its prediction error exceeds the current training RMSE by $2\;\%$. As the RMSE decreases during training, the threshold for accepting structures becomes lower, and the NNP starts accepting less important structures from the dataset. 

\textbf{Training data generation.} The structure generation for our workflow follows these steps:

\begin{enumerate}
  \item SQS cells relaxation 
  
We sample a representative subset of structures from the DFT relaxation trajectories generated for the $243$ SQS cells sampled by CE. The FPS algorithm based on DFT energies of ionic steps is utilized to remove highly similar structures, retaining approximately $70\;\%$ of the original trajectory. The counts of relaxation snapshots for each of the training compositions are listed in Table S1 under $N_{SQS\;strucs}$.
  \item Bulk ground states
  
For half of the training compositions, isotropic strain is applied to the bulk ground-state unit-cell structures, varying their lattice parameters by $\pm 10\;\%$. Subsequently, the NNP is used to relax the atomic positions of the strained structures while keeping the cell volume fixed. Specific structural snapshots are then chosen for self-consistent DFT calculations, employing a committee of two NNPs. These two NNPs differ slightly in configuration and the random seed used during the stochastic training process, as well as in the train/test dataset splitting \cite{singraber2019parallel}. Whenever large discrepancies between the NNPs' energy and force predictions are observed, the corresponding structures are selected for DFT calculations, leading to dataset updates.

Moreover, the NNP is utilized to perform NVT-ensemble MD simulations for $2 \times 2 \times 2$ supercells of the ground-state structures at three compositions close to the ground-state composition (Cr$_{0.667}$S), specifically, Cr$_{0.7}$S, Cr$_{0.625}$S, and Cr$_{0.5}$S. The MD trajectories encompass various temperatures ($300$ K, $600$ K, $900$ K) at isotropically strained volumes ($\pm 5\;\%$ and $\pm 10\;\%$ strain in lattice constant), in addition to the equilibrium volume. Similarly, snapshots are selected from the MD trajectories employing a two-NNP committee, ensuring that each pair of successively selected snapshots are separated by at least $100$ fs to minimize statistical correlations. The number of bulk ground-state snapshots generated using both the strain/relax and strain/MD approaches is recorded for the respective compositions in column $N_{{GS\;bulk}}$ of Table S1.
  \item Slab cuts of ground states 
  
We generate two-unit-cell-thick slab structures for the ground states at the three lowest-energy training compositions: Cr$_{0.7}$S, Cr$_{0.667}$S, and Cr$_{0.625}$S. Firstly, the bulk unit cell is shifted incrementally along a randomly chosen cell vector, then a vacuum layer larger than $15$ \AA{} is added in the direction of that cell vector to create 2D slabs with diverse surface terminations across the crystal. These slabs are subjected to $\pm 5\;\%$ strain around equilibrium and then fully relaxed using the NNP model. Structural snapshots are sampled from the relaxation trajectories using the NNP committee method, followed by self-consistent DFT calculations and dataset updates. The number of generated slab structures for each composition is recorded in column $N_{{GS\;slab}}$ of Table S1.
\end{enumerate}
Fig. S4 presents a reduced-dimensionality representation of the feature space spanned by the atomic environments in the above dataset, providing a means to evaluate structural similarities between the different structures comprising the dataset. The above DFT dataset comprised a total of $10,593$ structures. In addition, for subsections B and C in the Results, we augmented our training set with $380$ structures sampled by FPS from fixed-cell-volume relaxation trajectories of the vdW CrS$_2$ bulk unit cell. These structures were rendered at various isotropic compressive strains ranging from $0\;\%$ to $10\;\%$ along each lattice vector, with increments of $2\;\%$. This augmentation was done to particularly enhance the accuracy of NNP predictions at high compressive strains of CrS$_2$. 

\textbf{Simulated annealing.} The SA procedure employed for crystal structure optimization, as detailed in Results - Subsection B, commences with the system's initial equilibration at a high enough temperature (T = $21,000$ K), which enables a relatively unrestrained exploration of the configurational space. Subsequently, the system undergoes a gradual cooling process (following a geometric sequence) consisting of $50,000$ steps (which is found sufficient for our supercells to achieve energy convergence), culminating in the convergence toward a local minimum at a relatively low temperature (T = $0.01$ K). This cooling strategy facilitates occasional uphill moves in the early stages, prompting an exhaustive search for low-energy basins in the PES. To enhance the exploration of the low-energy crystal structures, we introduce a cyclic restart strategy, consisting of $12$ cycles, into our SA workflow. At the conclusion of each cycle, the temperature is reset to a value $1000$ K lower than the initial temperature, thereby inducing diverse cooling rates and promoting a more thorough exploration of the low-energy crystal structures. Each iteration during SA involves either the exchange of two random groups of vacancies and Cr atoms ($88\;\%$ of total steps) or the perturbation of lattice parameters of the simulation cell ($12\;\%$ of total steps) to maintain a zero-mean pressure. In the case of cell perturbations, the atomic positions are scaled with the stretched lattice to preserve fixed fractional coordinates, and the energies are computed using the NNP statically (without further relaxation of positions). The acceptance or rejection of these moves adheres to the Metropolis criteria for isobaric-isothermal (NPT) ensemble. Following this, the NNP model is employed to conduct full structural relaxations toward the neighboring local minima for the set of distinct configurations derived from the SA cycles. This endeavor aims to accurately discern the ground-state configuration amidst various alternative metastable structures.

\textbf{Vacancy diffusion Monte Carlo.} The MC simulations utilized for modeling Cr vacancy diffusion during non-vdW to vdW phase transformation in CrS$_2$ slabs, as discussed in Results - Subsection C, are conducted in the NPT ensemble, maintaining a temperature of $300$ K and zero absolute pressure in the $z$-direction. Starting with a randomized Cr/vacancy distribution in the Cr sublattice, each MC step involves either a perturbation in the slab thickness ($12\;\%$ of total steps) or a proposal a group of randomly chosen vacancies to move to any of their nearest neighbor (nn) sites in the Cr sublattice ($88\;\%$ of total steps). The nn list includes the surrounding $6$ Cr sites in a hexagonal arrangement within the same plane, as well as the analogous $7$ sites in each of the upper and lower Cr layers, as illustrated in Fig. \ref{fig:induced_vdw_gap} a). Each step of vacancy diffusion can be conceptualized as either remaining within the same layer or executing a vertical jump to an adjacent Cr layer, subsequently choosing to either stay at the new site or execute a horizontal movement to a nn in the new layer. This results in equal probabilities for movements in the $x$-$y$ plane and along the $z$-direction, ensuring unbiased vacancy transitions within the slab. 
During the MC simulation, we utilize an exponential growth function to adjust the number of permissible minimization steps, wherein the rate decreases in proportion to the MC steps. After conducting multiple tests with different rates and maximum values, we settled on the function $80 \times e^{(-\text{step}/10^5)}$ to achieve a nearly zero mean pressure with minimal fluctuations in the $z$-direction. Thus, this choice minimizes the forces on atoms while ensuring a steady-state equilibrium of the slab with the surrounding environment in the $z$-direction, along which the slab is exposed to vacuum. The MC simulation is terminated when a reduction of less than $1$ meV/atom in the cohesive energy of the slab is observed over a period of $10,000$ steps, resulting in a total of $210,000$ steps (see Fig.\ref{fig:induced_vdw_gap} c)).

\section{Results}

\subsection{Generalization of NNP for nonstoichiometric crystal structure optimization}

Crystal structure optimization in the context of nonstoichiometric solids typically involves two primary aspects: (i) configurational optimization, which involves determining the equilibrium arrangement of vacancies on the lattice sites, and (ii) local geometric optimization, which pertains to the local adjustment of structural parameters, such as atomic positions and cell dimensions. Different vacancy configurations can be visualized as distinct basins on the potential energy surface (PES), while the geometrically relaxed structures represent the local minima within these basins. While addressing both configurational and geometrical optimizations is crucial in global optimization problems, it is often reasonable to restrict this coupling to the low-energy basins within the PES \cite{gu2017prediction, oganov2011modern}. This approach is warranted by the constrained influence of local geometrical optimization on the energy of unfavorable configurations, attributed to the presence of barriers between the distinct basins. Thus, a configuration that hinders the formation of stable bonds is highly likely to persist as a high-energy structure even after undergoing geometrical optimization.

To evaluate the crystal structure prediction capability of our NNP model at different compositions, we initially assess its ability to rank on-lattice configurations with different vacancy arrangements across various unseen compositions. By decoupling off-lattice geometrical optimization from on-lattice configurational optimization (which is a strength of the CE method), we enable more meaningful comparisons between the NNP and CE models. Subsequently, we examine the NNP model's ability to accurately reproduce energies and forces for the lowest-energy configurations subjected to elastic strains and atomic perturbations from lattice sites. By successfully reproducing these energy and force profiles, we can validate the NNP model's effectiveness in fitting the PES around the local minima of low-energy basins, thus showcasing its competence in performing local geometric optimization.

\subsubsection{Configurational Optimization}

\begin{figure*}
\raggedright
\includegraphics[scale=0.24]{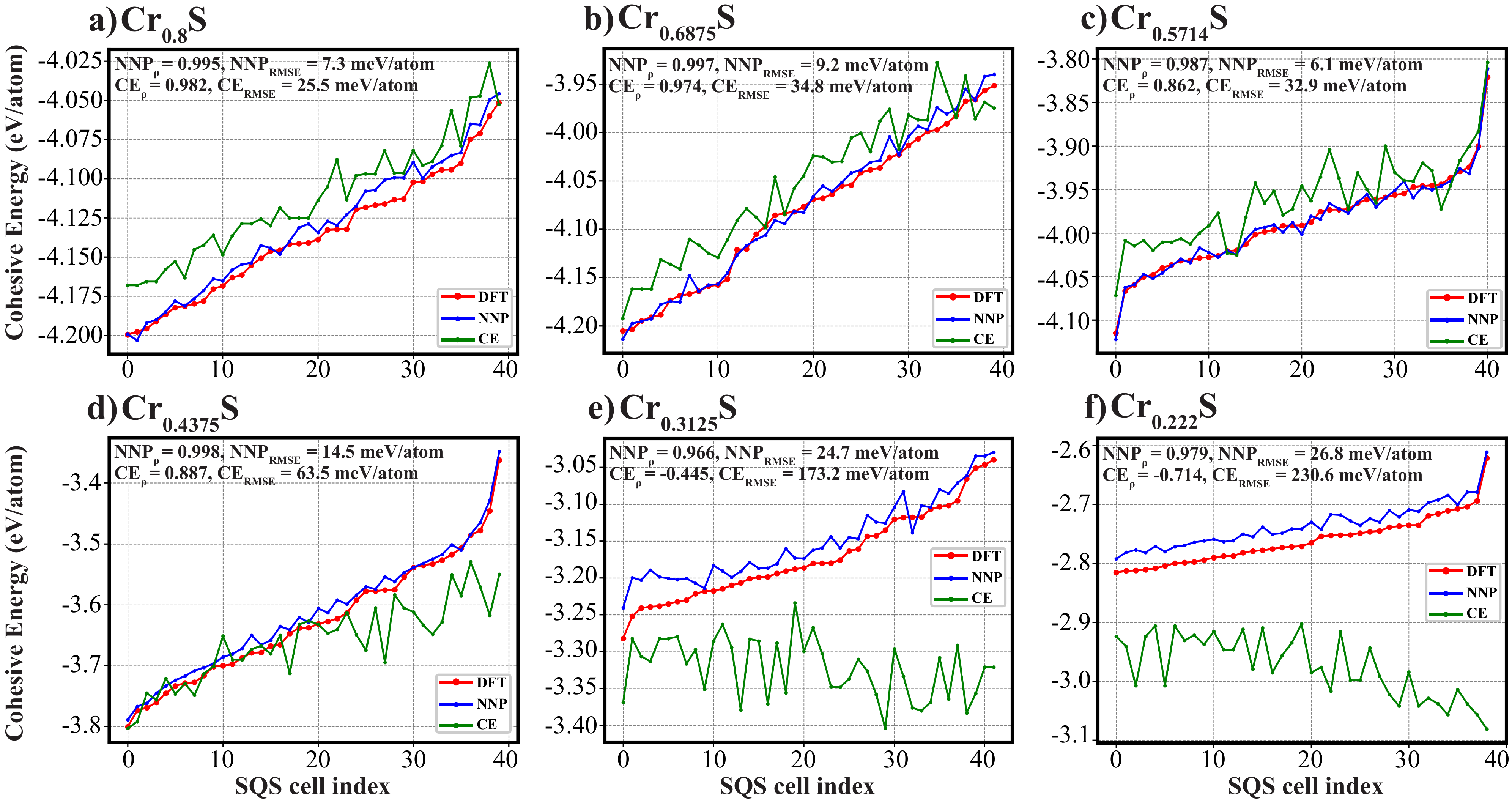}
\caption{\textbf{%Energy predictions for SQS cells %over total energy range
%at unseen compositions:
Evaluation of the NNP for configurational optimization at unseen compositions.} The ranking of around $40$ SQS cells, at six different unseen compositions, is demonstrated using CE, NNP, and DFT. The depicted ordering is based on the DFT energies, as observed from the monotonically rising DFT energy with cell index. The SQS cells span the entire energy range of the enumerated SQS cells ($< 48$ atoms per cell). }
\label{fig:ranking}
\end{figure*}

To minimize the cost associated with dataset generation during the training of the NNP model, we limited our consideration to small SQS cells with fewer than $20$ atoms during CE sampling. As a result, the composition space in the DFT dataset used for training/validating the NNP model was restricted to the $22$ compositions outlined in Table S1. In order to broaden the range of compositions accessible for assessing the NNP generalization, we increase the maximum number of allowable atoms per SQS cell from $20$ (as used in CE sampling) to $48$. This enables us to accommodate a wider spectrum of elemental composition proportions of Cr and S atoms. Subsequently, we select a new set of $18$ unseen vacancy compositions to validate the generalization of our model. To ensure a comprehensive assessment of the NNP model's ability to explore the entire configurational space for these unseen compositions, we opt for an exhaustive approach instead of employing any form of configuration sampling. This involves enumerating all possible SQS cells for the chosen compositions. The pool of enumerated SQS cells for the new $18$ compositions encompassed approximately $130,000$ symmetrically-distinct configurations. Next, we employed the NNP model to rank the enumerated SQS cells at each composition based on their cohesive energies. Initially, the energies of all SQS cells were statically computed without relaxation. Then, for the $300$ SQS cells with the lowest energies at each composition, a full relaxation procedure was implemented. This correction step aimed to rectify any potential inaccuracies in the ranking of the low-energy configurations caused by the initial constraint on geometrical degrees of freedom.

Among the numerous SQS cells available for each composition, we purposefully select a subset of around $40$ configurations. These configurations are chosen to span the entire energy range with approximately equal intervals. In order to ensure a fair comparison between the NNP and the CE models, we use here a second CE model fitted to the DFT energies of the SQS cells in the dataset, computed statically on-lattice without relaxation. This approach helps to eliminate any potential bias that may arise from the first CE model which was trained on DFT energies of fully relaxed SQS cells, and then utilizing it for predicting the energies of on-lattice configurations. The comparison of the NNP and the CE models against DFT is illustrated in Fig. \ref{fig:ranking}, focusing on six representative unseen compositions. The accuracy of each model is assessed using two key metrics: Spearman's rank-order correlation coefficient \cite{sheskin2003handbook}, denoted as $\rho$, which measures the degree of similarity in the energy-based ranking order of SQS cells, and RMSE, which quantifies the discrepancy between the predicted and actual energies.

As illustrated in Fig. \ref{fig:ranking}, the NNP demonstrates superior performance compared to CE, yielding RMSE values akin to those obtained during train/test evaluation (refer to Table S3), with all $\rho$ values exceeding $0.96$. Nevertheless, a slight reduction in performance is noted for Cr$_{(1-x)}$S configurations with vacancy concentration, $x$, above $\sim 0.55$ (as can be noticed from Fig. \ref{fig:ranking}-f). However, this can be deemed acceptable for such elevated vacancy concentrations which typically correspond to experimentally infeasible structures. Even if these structures are encountered in theoretical simulations, they are likely to be rejected due to their inherent instability. Conversely, CE exhibits a comparable performance to NNP, albeit to a lesser degree, for compositions that span relatively low energies; however, the performance deteriorates significantly for high-energy compositions, where the model entirely loses its ability to rank the different configurations for vacancy concentrations beyond $x \sim 0.55$. The notable advantage of the NNP over CE, despite both models being trained on the same configurational arrangements (i.e., the same SQS cells), highlights the greater flexibility of the NNP in effectively generalizing to diverse crystal structures with varying chemical occupations and compositions. 

\begin{figure*}
\centering
\includegraphics[scale=1.25]{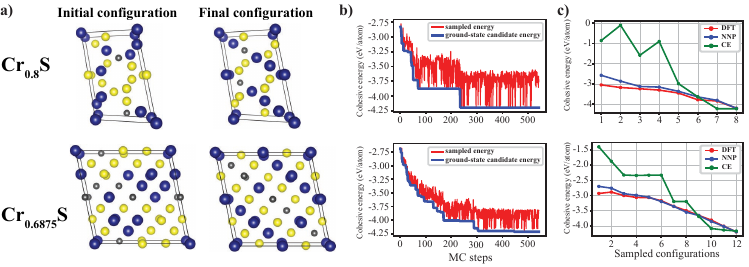}
\caption{\textbf{Generalization of the NNP for extrapolative-sublattice configurations.} a) The chemical order of the ground-state configuration is initially randomized for two unseen stoichiometries (Cr$_{0.8}$S and Cr$_{0.6875}$S), allowing any species (Cr, S, or vacancy) to occupy any site (not constrained to the training Cr/vacancy and S sublattice partitioning). b) NNP is utilized to perform short MC simulations involving atomic swaps until the ground state is retrieved. c) Different configurations are sampled from the MC trajectory at nearly equal energy intervals and subsequently validated against CE and DFT.}
\label{fig:MC}
\end{figure*}

Nonstoichiometric systems are typically treated using sublattice models. Accordingly, we partitioned the parent stoichiometric Cr$_2$S$_2$ lattice into two sublattices: a sublattice of active sites (Cr, vacancy) and another sublattice of inactive S sites, in accordance with experimental observations. All the SQS cells employed in training and validation were constrained within this partitioning. Despite the rationality of this sublattice approach, it is of interest to investigate the ability of the NNP to generalize to arbitrary chemical occupations, where each species (Cr, vacancy, S) can occupy any site in the cell. In this context, we select the ground-state SQS cells identified for two unseen compositions, Cr$_{0.8}$S and Cr$_{0.6875}$S, and then randomize the chemical occupation in each cell to generate initial disordered configurations, as depicted in Fig. \ref{fig:MC} a). Subsequently, we conduct simple MC simulations using the NNP. In each step, a MC swap of two sites of different species is randomly suggested, and a greedy algorithm is employed to accept or reject the move. The NNP model successfully restores the ground-state configurations, as shown in Fig. \ref{fig:MC} b), demonstrating its accurate ranking capability by rejecting highly energetic configurations until reaching the true ground state. Furthermore, we sample a number of configurations from the MC trajectories at nearly equal energy intervals and calculate their DFT energies. From Fig. \ref{fig:MC} c), it is evident that the performance of the NNP retains its high quality even for highly randomized configurations with cohesive energies above $\sim -4$ eV/atom, which is roughly the highest energy observed for the SQS cells of the two considered compositions (Cr$_{0.8}$S and Cr$_{0.6875}$S) when S atoms were considered inactive (see Fig. \ref{fig:ranking}). Only in the vicinity of the highest energy configurations the deviation of NNP-predicted energies starts to increase; however, the ranking predictions remain in very good agreement with DFT throughout the entire energy range. On the other hand, CE successfully identifies the random configurations as unfavorable with high-energy predictions, albeit with considerably higher error values compared to the NNP. The overall remarkable capability of the NNP to accurately rank unseen configurations across a broad spectrum of energies and diverse compositions underscores its robustness as a promising tool to drive MC simulations and generate more reliable trajectories for atomic diffusion processes.

\subsubsection{Local geometrical Optimization}

The preceding part elucidated the ability of the NNP to exhibit generalization across various atomic configurations, covering the space of SQS cells across a wide range of compositions. In this part, we analyze the efficacy of NNP to approximate the PES in the vicinity of local minima corresponding to ground-state structures for different compositions. Specifically, we focus on the lowest-energy SQS cells at $18$ unseen compositions, as determined by NNP from the enumerated SQS cells.  

\begin{figure*}
\centering
\includegraphics[scale=0.5]{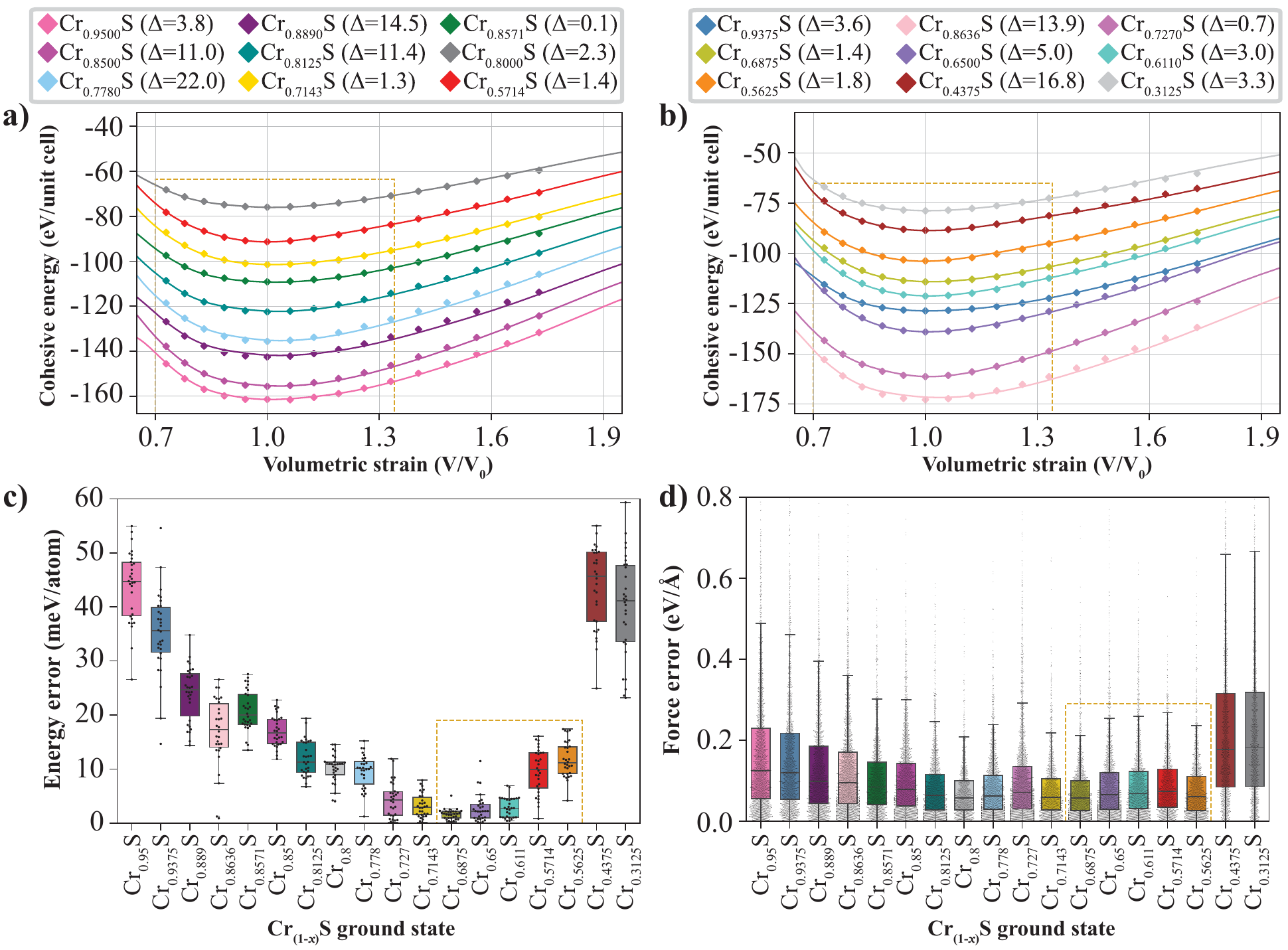}
\caption{\textbf{Generalization of the NNP for elastic and vibrational perturbations of ground-state structures at unseen compositions.} a) and b) present the cohesive energy versus volumetric strain curves as predicted by NNP (solid lines) compared to DFT reference values (discrete markers). The structures are divided between panels (a) and (b) only for clarity purposes. Numerical error estimations of NNP predictions are provided by the listed $\Delta$ values (in meV/atom) for each structure. The dashed yellow boxes in (a) and (b) indicate the volumetric strain range that NNP witnessed for some training structures. c) and d) depict box plots for NNP energy and force error distributions based on $30$ snapshots sampled from $500$ K MD simulations at $3$ ps intervals. The rectangular box indicates the interquartile range (IQR), while the line within the box represents the median. The dashed yellow boxes in (c) and (d) specify the composition range over which NNP was trained on high-temperature configurations of ground-state structures (specifically for Cr$_{0.7}$S, Cr$_{0.625}$S, and Cr$_{0.5}$S).}
\label{fig:EOS}
\end{figure*}

To evaluate the performance of NNP at various bond lengths that significantly deviate from equilibrium, we subject the relaxed ground-state configurations to isotropic strain ranging from $-10\;\%$ to $+20\;\%$ along each lattice vector, with an increment of $2\;\%$. Panels a) and b) in Fig. \ref{fig:EOS}, illustrate the NNP predictions of cohesive energies at these volumetric strains for the ground-state structures associated with the $18$ unseen compositions. The NNP predictions are computed over a fine-grained grid of strains, visually represented by solid lines in the figure. The discrete markers, on the other hand, represent the reference values obtained from DFT, revealing a remarkable concurrence with the NNP predictions across a wide range of test volumes and varying compositions. It is noteworthy that the error in the NNP predictions remains relatively uniform throughout the strain range, albeit with a slight increase observed at extreme strains. In order to provide a quantitative assessment of the accuracy achieved by the NNP, we utilize a measure, $\Delta$, which is commonly employed to compare equations of state derived from different models \cite{zuo2020performance, lejaeghere2014error, csonka2009assessing}.
\begin{equation}
\Delta=\frac{\int_{(0.9)^3 V_0}^{(1.2)^3 V_0} |E^{\text{NNP}}(V)-E^{\text{DFT}}(V)| \;d V}{((1.2)^3-(0.9)^3) V_0}
\end{equation}
where the $E^{\text{NNP}}(V)$ and $E^{\text{DFT}}(V)$ functions are obtained by fitting a 6th-order polynomial to the corresponding data, $V_0$ denotes the equilibrium cell volume, and the $0.9$ and $1.2$ values correspond to the $-10\;\%$ and $+20\;\%$ applied strain to each lattice vector. The $\Delta$ value for each composition is outlined (in meV/atom) at the top of Fig. \ref{fig:EOS} a) and b). It can be observed that two-thirds of the $\Delta$ values are below the estimated RMSE of the NNP (refer to Table S3), while the largest error of $22.0$ meV/atom is about 2.5 times that value. It should also be noted that the NNP was trained on strained ground-state structures associated with training compositions (referred to as $N_{GS\;bulk}$ in Table S1), where the volumetric strain during training was limited to the range from $0.73$ to $1.33$ times the equilibrium volumes. Nevertheless, NNP still matches DFT values for larger strains up to $1.73$ of the equilibrium volumes. However, we observe larger deviations when attempting to extrapolate to volumetric strains smaller than $0.7$. This can be attributed to the relatively higher forces associated with such large compression strains, which the NNP model might not have been sufficiently trained to accurately capture and represent.

Next, we focus on the capacity of NNP to accurately replicate the energies and forces obtained from DFT during MD simulations. This aspect is of considerable importance as it directly influences the model's proficiency in accurately relaxing atomic positions and predicting vibrational properties. In this retrospect, we run NNP-driven MD simulations for the relaxed unit cells of the $18$ test structures at $500$ K with a $0.5$ fs timestep. Then, we sample $30$ snapshots from the MD trajectory, and perform self-consistent DFT calculations to verify the energies and forces on the sampled structures. To mitigate correlations between the sampled structures, we adjust the time interval between each two sampled snapshots to $3$ ps.  Panels c) and d) in Fig. \ref{fig:EOS} demonstrate the error distribution of energies and forces across ground-state structures of the $18$ compositions. It is worth noting that the NNP has been only trained on high-temperature MD configurations for Cr$_{0.7}$S, Cr$_{0.625}$S, and Cr$_{0.5}$S ground states. Despite this, we observe that the median energy error is below $20$ meV/atom for structures ranging from Cr$_{0.8636}$S to Cr$_{0.5625}$S. A similar behavior is observed for force errors, where we can see that the medians of force errors are below $0.2$ eV/\AA{} for all compositions, and aside from a few outliers, the full distributions of force errors are smaller than $0.3$ eV/\AA{} in the composition range from Cr$_{0.8636}$S to Cr$_{0.5625}$S. This reveals the interpolation ability of the NNP, allowing it to accurately approximate the vibrational properties of ground-state structures at unseen compositions. Although the NNP exhibits a good extrapolation ability outside the training composition range, even beyond the compositions used in the high-temperature MD training (Cr$_{0.7}$S, Cr$_{0.625}$S, and Cr$_{0.5}$S), we speculate that this behavior is primarily captured from training on the relaxation structures of nearby training compositions. Consequently, we observe a gradual increase in energy and force errors as we deviate away from the composition range that comprised the high-temperature MD training (denoted by the yellow box in Fig. \ref{fig:EOS}). This occurs because, for these compositions, the energy and force predictions are solely reliant on similarities to the relaxation structures sampled at proximate compositions. To provide similarity measures between the unseen structures utilized for assessing the configurational and geometrical generalization of the NNP model, we present in Fig. S5 a comparison between the reduced-dimensionality representation of the feature space spanned by the atomic environments in these structures and that of the DFT dataset.

\subsection{Crystal structure predictions for Cr$_{(1-x)}$S}

The characterization of material properties, encompassing electronic band structure, density of states, and magnetic characteristics, exhibits a profound reliance on the fundamental crystalline framework (comprising atomic occupations and positions). It is imperative to ensure a precise representation of the input crystal structures in order to enhance the predictive capacity of DFT in property prediction, as any deviations from realistic structures may introduce uncertainty into property predictions.

\begin{figure*}
\centering
\includegraphics[scale=0.126]{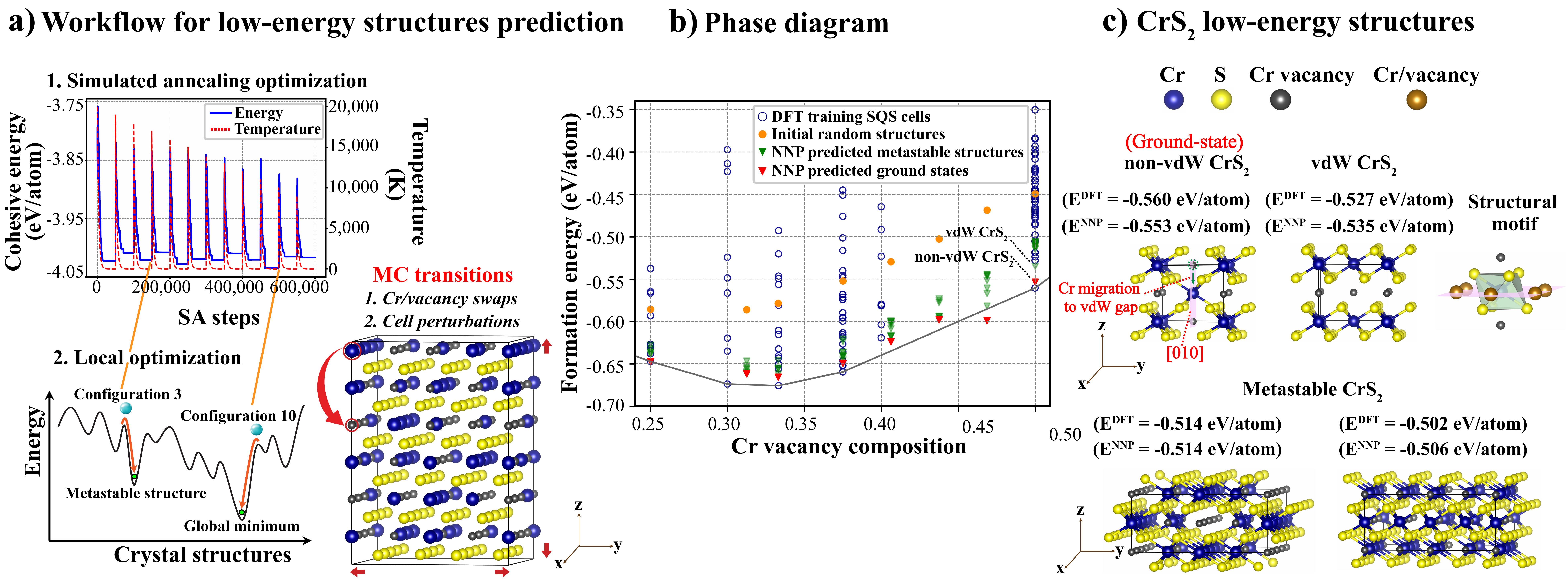}
\caption{\textbf{Simulated annealing prediction of Cr$_{(1-x)}$S stable structures:} a) depicts the NNP-driven crystal structure prediction workflow: 1. Twelve-cycle simulated annealing (SA) with Cr/vacancy swaps and cell perturbations. The represented structure illustrates the allowed MC transitions (red arrows) in the adopted supercell during SA. 2. Local geometry relaxations of the candidate configurations to identify global minimum from metastable structures. b) illustrates the phase diagram of Cr$_{(1-x)}$S in the range $x = [0.25, 0.5]$ along with the identified ground-states and metastable structures. c) exhibits the refined cells of the ground-state and few metastable structures for the CrS$_2$ phase.}
\label{fig:simulated_annealing}
\end{figure*}

%In this subsection, we implement the NNP model to derive accurate stable structural representations of Cr$_{(1-x)}$S phases across different compositions.
In this subsection, we implement the NNP model to accurately explore the low-energy Cr$_{(1-x)}$S crystal structures across different compositions. To achieve this, we utilize the SA algorithm, a well-established and robust optimization technique in the field of crystal structure optimization \cite{oganov2011modern, doll2008structure, pannetier1990prediction}. To enhance the exploration of the low-energy crystal structures, we incorporate a cyclic restart strategy (comprising $12$ cycles) within our SA workflow, yielding an ensemble of up to 12 distinct Cr/vacancy configurations, ensuring a trade-off between diversity and computational efficiency. The SA procedure is illustrated in Fig. \ref{fig:simulated_annealing} a), and additional details can be found in the "Methods" section.

In this study, we limit our SA work to identify the low-energy Cr$_{(1-x)}$S structures in the vacancy concentration range $(0.25 \leq x \leq 0.5)$, aligning with experimental interest. We choose $8$ different compositions to sample this range. To encompass a diverse set of compositions, we employ $4 \times 2\sqrt{3} \times 3$ supercells, which are derived from the orthorhombic ($Pnnm$) Cr$_2$S$_2$ conventional unit cell. This supercell size of lattice parameters in the range of $12.0$ \AA\;to $16.0$ \AA\;allows us to strike a balance between mitigating finite-size errors and ensuring appropriately sized unit cells for future DFT investigations. Figure \ref{fig:simulated_annealing} b) illustrates the comprehensive phase diagram of Cr$_{(1-x)}$S across the range of compositions under investigation. The diagram presents formation energies, computed with reference to the most thermodynamically stable phases of Cr and S elements as available in the Materials Project database \cite{jain2013commentary}, namely mp-90 (Cr in BCC lattice) and mp-96 (S in P2/c space group), respectively. The depicted energies encompass the NNP-predicted values for both the initial random configurations at  $0^{th}$ step of each SA process and the distinct low-energy structures, including both ground-states and metastable configurations, obtained for the $8$ considered compositions. Additionally, the diagram includes the DFT formation energies of the SQS cells present in the NNP training set within the same composition range. The applied workflow effectively identified the ground-state ordered structures corresponding to all known phases in the considered composition range, namely Cr$_{0.75}$S (Cr$_3$S$_4$) (mp-964), Cr$_{0.667}$S (Cr$_2$S$_3$) (mp-555569), and Cr$_{0.375}$S (Cr$_5$S$_8$) (mp-1181961) (see Fig. S2 and Fig. S3). This validation confirms the precision and reliability of both the NNP model and the SA workflow in predicting ground-state structures.  

Notably, a consistent observation within the investigated composition range reveals a pronounced preference for ground-state structures that feature complete CrS$_2$ layers intercalated by Cr-deficient layers (see the predicted ground states and the layer occupation distributions in Fig. S2). This intriguing finding indicates a tendency for vacancies to aggregate forming alternating (CrS$_2$ and Cr-deficient) layers rather than being uniformly distributed across all layers. Hence, the presence of $\delta$ excess intercalant Cr atoms above the Cr/S = 1/2 ratio (i.e., Cr$_{(1+\delta)}$S$_2$) occupying the vdW gaps manifests as a critical factor in stabilizing CrS$_2$ layers, preventing Cr migration from them into the vdW gaps, even for small $\delta$ values of down to approximately $0.03$, as observed for the Cr$_{0.53125}$S phase. Upon reaching the Cr$_{0.5}$S composition, the energy cost associated with forming CrS$_2$ layers and empty vdW gaps in-between becomes higher, leading to a compelling phenomenon where half of the Cr atoms preferentially migrate to the vdW gaps, forming a non-vdW CrS$_2$ structure. This manifests that CrS$_2$ does not prefer the vdW nature as already observed in experiments \cite{song2019soft, stiles2023unlocking, song2021properties, song2021kinetics}. Figure \ref{fig:simulated_annealing} c) presents both the ground-state non-vdW CrS$_2$ and the subsequent stable phase, the pure vdW CrS$_2$. The non-vdW CrS$_2$ configuration can be conceptualized as a transformation from the vdW CrS$_2$, wherein $50\;\%$ of the Cr atoms undergo orderly shifts toward the vdW gap along the $[010]$ direction, resulting in an AB stacking pattern, distinct from the AA stacking seen in vdW CrS$_2$. The DFT formation energy of non-vdW CrS$_2$ is lower by $33$ meV/atom, underscoring the non-vdW character of Cr-S systems. A more precise calculation, incorporating vibrational free energy contributions associated with phonons, is illustrated in Fig. S6. The results indicate that the free energy of the non-vdW CrS$_2$ phase is predicted to be lower for temperatures up to around 465 K, affirming its higher stability compared to the vdW phase. Moreover, it should be noted that the reported DFT results are obtained using the standard PBE functional without a vdW correction term. However, including a vdW correction is observed to increase the binding of the non-vdW phase more than the vdW one, thereby predicting larger relative stability of the non-vdW phase.

Nevertheless, it is imperative to recognize that solely prioritizing the acquisition of the ground-state (global minimum) structure may be insufficient in faithfully representing the material's experimental reality. This is due to the presence of alternative low-energy local minima in the PES, which correspond to metastable phases capable of being adopted by the material during the synthesis process or through thermal excitation. This becomes particularly significant in the case of 2D non-vdW materials, such as Cr-based 2D materials, where the interlayer bonding forces between atoms are typically stronger than vdW interactions. %Consequently, atoms in these materials tend to exhibit a higher propensity to migrate from the Cr-rich layers to the Cr-deficient layers. 
As a consequence, rapid and uncontrolled interlayer diffusion may occur from the Cr-rich layers to the Cr-deficient layers, restricting the scope of lateral re-arrangement of Cr atoms during synthesis. This phenomenon ultimately leads to the stabilization of metastable structures with smaller ranges of order, thereby hindering the material from attaining its global minimum long-range ordered crystal structure. 

For instance, during the soft chemical synthesis of CrS$_2$ slabs derived from bulk NaCrS$_2$ through proton exchange, the deintercalation of Na induces a notable degree of disorder within the Cr sublattice \cite{song2019soft, song2021properties}. This is evident as the Cr atoms undergo rapid and uncontrolled transitions from the CrS$_2$ (Cr-rich) layers to the vacant Na (Cr-empty) sites. As a result, a metastable CrS$_2$ structure forms, where both Cr atoms and vacancies are randomly distributed throughout the Cr-sublattice, forming a 3D chemically bonded arrangement. Figure \ref{fig:simulated_annealing} c) represents refined unit cells of two metastable CrS$_2$ structures observed during our SA. The metastable structures are not perfectly ordered as the non-vdW or vdW structures, however, the prominent structural motif in these two ordered structures can still be observed in the two metastable structures with higher formation energies. In particular, we can notice that each Cr atom consistently exhibits a preference for being neighbored by vacant sites in the adjacent top/bottom Cr layers, while its lateral surroundings entail a disordered disposition of both Cr atoms and vacancies. This configuration more accurately reflects both the non-vdW and disordered characteristics observed in the experimentally synthesized 2D CrS$_2$ slabs, where roughly half of the Cr atoms undergo random interlayer jumps from the CrS$_2$ layers to the vdW gaps, resulting in the creation of randomly scattered Cr sites neighbored by vacant sites in the adjacent top/bottom Cr layers. As a consequence, these metastable structures manifest as better representative supercells enriched with more realistic motifs for DFT property predictions. This behavior is not exclusive to CrS$_2$, as it is also experimentally observed in other phases. For instance, the ground-state ordered phase of Cr$_2$S$_3$ displays a structural configuration akin to CrS$_2$-Cr$_{1/3}$-CrS$_2$, wherein Cr$_{1/3}$ forms Cr lines extended along the armchair direction, separated by two vacant lines in either ABC or AB stacking \cite{liu2022atomic, chu2019sub, cui2020controlled}.  However, structural characterization of CVD-grown Cr$_2$S$_3$ slabs reveal a pronounced degree of disorder in the Cr-deficient layers, leading to the crystal being characterized as a CrS$_2$-disordered Cr$_{1/3}$-CrS$_2$ \cite{liu2022atomic}. The same thing applies for Cr$_3$S$_4$, being characterized as a CrS$_2$-disordered Cr$_{1/2}$-CrS$_2$ \cite{liu2022atomic}.

\subsection{Strain-induced vdW gaps in non-vdW CrS$_2$ slabs}

In the preceding sections, we discussed that Cr atoms within Cr-S structures exhibit a high propensity to migrate from the CrX$_2$ layers to partially occupy the vdW gaps \cite{song2019soft, stiles2023unlocking, song2021properties, song2021kinetics}, thereby establishing strong 3D chemical bonds within the crystal lattice. The presence of such chemical bonds between adjacent layers in these CrX$_2$ materials poses challenges in achieving precise thickness control during the exfoliation process, limiting the attainment of ultra-thin sheets down to monolayer thickness. For the case of CrS$_2$ exfoliation, the deintercalation of Na from NaCrS$_2$ followed by LPE typically yields sheets with a minimum thickness of approximately $2$-$3$ nm \cite{song2019soft, song2021kinetics, song2021properties}. Furthermore, our findings in Results - Subsection B demonstrate that the ground-state structure of CrS$_2$ favors an isotropic distribution of vacancies throughout the layers of bulk crystal rather than forming a vdW CrS$_2$ structure. In this part, we show that lateral compressive strain is a viable method for inducing the reconstruction of vdW gaps between the CrS$_2$ layers. This effect leads to a reduction in the interlayer interactions, thereby facilitating the exfoliation process of these materials. A similar phenomenon has been experimentally observed in other materials, such as germanium telluride, where lateral compression was found to induce the opening of vdW gaps \cite{yu2022strain}.

\begin{figure*}
\centering
\includegraphics[scale=0.118]{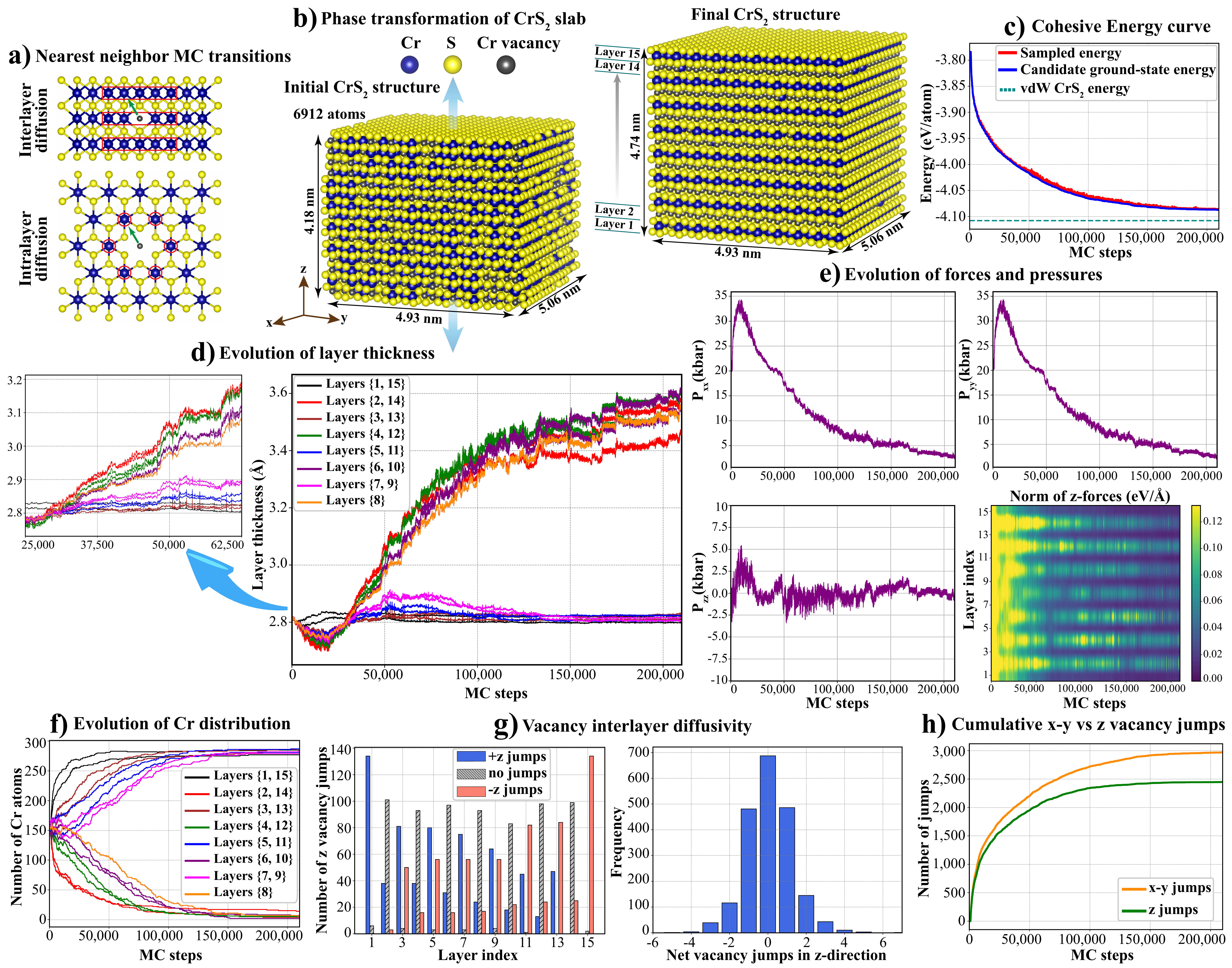}
\caption{\textbf{Strain-induced vdW gaps in CrS$_2$ slabs.} a) illustrates the nn vacancy transitions implemented during the MC simulation (a vacancy can jump to any of the $6$ surrounding sites identified by the red circles in the same plane, or to any of the $7$ nn sites in the adjacent top or bottom Cr layers). b) shows the initial random configuration along with the final configuration resulting from atomic rearrangements induced by lateral compressive strain. Layer numbering is indicated in the figure. c) shows the reduction in cohesive energy of the slab during the MC simulation. d) displays the thicknesses evolution of the $15$ layers over MC steps. e) illustrates the variation in pressure tensor diagonal components besides the norm of $z$-forces on Cr atoms for each layer in the slab. f) displays the evolution of the number of Cr atoms across the layers over MC steps. g) summarizes the count and direction of interlayer vacancy migrations from each layer (shown in the bar plot), along with the frequencies of net interlayer jumps of all vacancies from their initial layers (shown in the histogram). h) shows the cumulative numbers of interlayer and intralayer transitions made by all vacancies from their initial positions in the slab.}
\label{fig:induced_vdw_gap}
\end{figure*}

In this retrospect, we consider a symmetric CrS$_2$ slab with S termination on both sides. The slab consists of a $16 \times 9\sqrt{3} \times 8$ supercell of the orthorhombic ($Pnnm$) conventional unit cell (6912 atoms per simulation cell), with periodic boundary conditions applied in the $x$-$y$ directions and a $30$ \AA{} vacuum layer in the $z$-direction. The $x$-$y$ lattice constants are constrained to approximately $5.06$ nm $\times$ $4.93$ nm, while the thickness in the $z$ direction is allowed to freely relax starting from an initial value of $4.18$ nm. These $x$-$y$ lattice constants correspond to a lateral strain of around $3.5\;\%$ compared to the free-standing CrS$_2$ monolayer with a lattice constant of around $0.329$ nm. This strain value is approximately half of the strain observed in CVD-grown vdW CrS$_2$ slabs on f-mica substrate \cite{xiao2022van}, where the lattice mismatch is approximately $7\;\%$. We consider the average strain value in the CVD-grown CrS$_2$ slabs, assuming a linear variation of strain from $7\;\%$ to zero as we move away from the substrate. 

Starting from a non-vdW CrS$_2$ slab with a randomized Cr/vacancy distribution in the Cr sublattice, we utilize the NNP to conduct a vacancy diffusion MC simulation within the NPT ensemble, at $300$ K and zero absolute pressure in the $z$-direction. Throughout the MC steps, Cr atoms/vacancies undergo intralayer and interlayer diffusions to their nearest neighbors (see Fig. \ref{fig:induced_vdw_gap} a)), leading to a gradual transformation toward a quasi-vdW CrS$_2$ phase. Further details of the MC simulation procedure are elucidated in the "Methods" section. 

Figure \ref{fig:induced_vdw_gap} b) illustrates the initial and final structures of the MC simulation. Initially, the slab contains a nearly isotropic distribution of Cr atoms among the $15$ Cr layers (layer ordering is depicted in the figure). However, the application of compressive strain in the $x$-$y$ plane is shown to induce an ultimate alternating distribution of Cr atoms along the $z$-direction. Cr atoms undergo diffusion in a manner that results in the formation of almost complete vdW gaps, with approximately $97.8\;\%$ of the Cr atoms occupying the sites characteristic of a pure vdW CrS$_2$ phase. This behavior originates from the stresses induced by the compressive strain in the $x$-$y$ plane. As the slab is allowed to relax in the $z$-direction, it attempts to relieve the pressure through Poisson's effect, leading to expansion in the $z$-direction as seen in Fig. \ref{fig:induced_vdw_gap} b), where we can see that the slab experiences an expansion of $0.56$ nm in the $z$-direction. 

Figure \ref{fig:induced_vdw_gap} d) depicts the thickness evolution of the $15$ layers of the CrS$_2$ slab during the MC simulation. Initially, there is a transient period for approximately $10,000$ steps during which all inner layers undergo a reduction in thickness due to the inner pressure build-up in response to the externally imposed strain (see Fig. \ref{fig:induced_vdw_gap} e)), except for the surface layers (layers $\{1, 15\}$), which do not experience the same internal stresses owing to their direct exposure to vacuum. After this transient period, notable observations can be made from the data presented in Fig. \ref{fig:induced_vdw_gap} d): 

(i) All layers attempt to increase their thickness to alleviate lateral stresses. However, it is important to note that the level of resistance to this expansion varies among the layers, resulting in differing degrees of expansion for each layer. Specifically, the two surface layers (layers $\{1, 15\}$) exhibit the smallest degree of expansion. This is attributed to the net pulling force they experience toward the center of the slab due to their exposure to vacuum, which counteracts the slab's inherent tendency to expand for pressure release. As a result, these layers maintain their attained equilibrium thickness, preserving their vertical S-Cr-S bond lengths at an equilibrium state. Thus, layers $\{1, 15\}$ act as favorable sites for accommodating additional Cr atoms, resulting in the migration of vacancies away from them. 

(ii) In contrast, the neighboring inner layers $\{2, 14\}$ encounter significantly smaller net pulling forces, resulting in diminished resistance to axial expansion. Consequently, layers $\{2, 14\}$ undergo  more substantial increase in thickness, causing the S-Cr-S bond lengths in these layers to stretch beyond the equilibrium value. As a result, these layers become higher-energy sites for Cr atoms. Accordingly, Cr atoms start to migrate from layers $\{2, 14\}$ to the neighboring layers. The cumulative migration of Cr atoms from layers $\{2, 14\}$ results in the formation of low-density Cr environments adjacent to layers $\{3, 13\}$, inducing a net inward pulling force on the Cr atoms in layers $\{3, 13\}$, akin to the situation observed in layers $\{1, 15\}$. 

(iii) As a consequence, layers $\{3, 13\}$ exhibit higher resistance to expansion, thereby facilitating the migration of vacancies and displaying a stronger preference for accommodating Cr atoms. This phenomenon progresses toward the middle layers of the slab, leading to the formation of an alternating pattern characterized by lower-energy (Cr-rich) layers and higher-energy (Cr-deficient) layers. In Fig. \ref{fig:induced_vdw_gap} e), we can observe that the norm of z-forces acting on the even (Cr-deficient) layers exhibit higher values in comparison to the odd (Cr-rich) layers, thereby confirming the emergence of the alternating layered pattern. 

(iv) Additionally, Fig. \ref{fig:induced_vdw_gap} d) reveals how the layers' expansion behavior begins to bifurcate after approximately $30,000$ steps, leading to the formation of two distinct groups: one with increased thickness for the Cr-deficient layers and another with smaller thickness for the Cr-rich layers. This transition signifies the opening of vdW gaps transforming the initial strong 3D covalent bonding to weak interlayer vdW bonding between the covalently bonded CrS$_2$ layers. In other words, our findings detail the transformation of a non-vdW CrS$_2$ slab to form vdW CrS$_2$ layers.

Figure \ref{fig:induced_vdw_gap} f) illustrates the distribution of Cr atoms across the layers during the MC simulation. We can observe that the number of Cr atoms gradually decreases within the group of even (Cr-deficient/vdW gap) layers and increases within the odd (Cr-rich/CrS$_2$) layers where site energies are comparatively lower. More interestingly, the amplitude of diffusion rate (i.e., slope of the Cr distribution curve) exhibits a gradual increase as we approach the free surface of the slab at early stages of MC iterations. This reveals that Cr-rich/Cr-deficient alternating layer formation initially occurs at the free surfaces and propagates toward the middle layers of the slab. This point is further expounded in Fig. \ref{fig:induced_vdw_gap} g), wherein the extent of diffusivity is assessed through quantification of the number of vertical jumps made by vacancies from their initial positions. We can notice that the number of net vertical vacancy jumps is highest for the surface layers and decreases progressively toward the central layers of the slab. Notably, the even layers demonstrate a greater frequency of zero vacancy vertical displacements, owing to the higher-energy sites in these Cr-deficient layers that are amenable to the occupation of vacancies rather than Cr atoms. Furthermore, a symmetrical distribution of vertical vacancy jumps is observed in both the +$z$ and -$z$ directions, which indicates an unbiased diffusion along the $z$ axis, as one would naturally anticipate.

The vertical diffusion of vacancies in this process exhibits a localized nature, signifying a limited diffusion length. The histogram in Fig. \ref{fig:induced_vdw_gap} g) reveals that approximately $34\;\%$ of vacancies remain confined within their respective layers, while about $48\;\%$ diffuse to the first nn layer, $13\;\%$ move to the second nn layer, and only around $5\;\%$ diffuse beyond the second nn layer. This observation indicates a reduced likelihood of long-range interlayer jumps. Moreover, since the vertical displacement of vacancies is restricted to specific atomic neighborhoods, this is expected to lead to a higher occurrence of intralayer vacancy transitions. These intralayer transitions allow vacancies to explore various sites within their respective layers, potentially facilitating rearrangements that promote interlayer diffusion. Interestingly, the cumulative counts of intralayer and interlayer transitions, as depicted in Fig. \ref{fig:induced_vdw_gap} h), show comparable probabilities at initial MC steps, eventually reaching a plateau with an average of only $\sim 1.21$ horizontal vacancy jumps for each vertical jump, which further proves the limited diffusion length associated with intralayer transitions as well. These findings offer evidence that the prevailing diffusion process is of a short-ranged nature, characterized by movements occurring in immediate proximity to vacancies and their adjacent layers. As a consequence, this localized diffusion mechanism is expected to facilitate the transition from a non-vdW configuration to a vdW layered arrangement, owing to the reduced number of barriers that need to be traversed. These findings hold significant implications for the exfoliation of clean monolayer/few-layer 2D materials from non-vdW crystals through strain engineering, thereby unlocking new prospects for device-level applications and large-scale production of 2D materials.

%They also pave the way for exploring device-level applications and mass production of 2D vdW materials through chemical exfoliation methods.

%Overall, this analysis suggests that the initial lateral compressive strain can serve as a catalyst for the phase transition from a non-vdW structure to a vdW structure, allowing for the release of internal pressure via Poisson's axial expansion. 

\section{Discussion}

In this study, we presented a systematic approach for developing a highly efficient and accurate neural network potential to model nonstoichiometric environments prevalent during the chemical exfoliation of non-vdW systems, using Cr$_{(1-x)}$S as a case study, encompassing the entire range of vacancy concentrations in the Cr sublattice. We extensively evaluated the generalizability of the NNP model by testing its configurational and local geometrical optimization capabilities across various unseen compositions. 

In the domain of on-lattice configurational optimization, where the CE method is traditionally proficient, we have shown the NNP's superiority over CE, achieving significantly lower energy errors with respect to DFT. Notably, as we move from ground-state compositions of Cr$_{(1-x)}$S, concentrated around (Cr$_{0.667}$S), toward higher nonstoichiometric phases (especially at high vacancy concentrations), the accuracy gap between the NNP and CE models becomes more apparent in on-lattice configurational optimization. This observation highlights NNP's enhanced capacity to generalize to higher-energy basins in the PES, even without explicit training on similar configurations. In contrast, CE exhibits weaker generalization in high-energy regions unless specifically trained with a broader dataset encompassing both low-energy and high-energy configurations. A recent study supports this claim and demonstrates how training CE on a larger, diverse training set of configurations generated by MLPs can improve its performance and result in more robust thermodynamic simulations \cite{xie2022machine}. 

Although the NNP dataset contains more data points than the CE dataset, we should note that the configurations (atomic arrangements of Cr, S, and vacancies) in both datasets are the same. This is because the excess data points in the NNP dataset consist of merely snapshots from strain/MD or strain/relaxation trajectories of the same SQS cells that comprise the CE dataset. This argues that it is not the larger dataset that is responsible for the better performance but rather the higher flexibility of NNP, which provides better generalization. Indeed, the higher flexibility of the NNP does come with the trade-off of a more complex model compared to CE, with a much larger basis set size, as outlined in Tables S2 and S3. This necessitates the NNP to acquire a substantially larger dataset to train this expanded basis set. However, we note that around $80\;\%$ of the dataset provided for NNP was sampled from the non-deployed DFT relaxation calculations performed already to train CE. This suggests that instead of conventionally having CE as a final model for a nonstoichiometric system, it can be regarded as a first-step dataset-sampling model for the more accurate NNP model that will be trained mostly on the idle DFT dataset generated originally to fit CE. 

Furthermore, we have also shown that the robust generalization of NNP to high-energy configurations at unseen compositions persists even when extrapolating to sublattice configurations beyond the scope of the training data. Despite our dataset consisting exclusively of two-sublattice configurations (inactive S sublattice and active Cr/vacancy sublattice), the NNP accurately predicts configurations where Cr, S, and vacancies are allowed to occupy any lattice site.

Moreover, we meticulously examined the NNP model's accuracy in approximating the PES around the local minima corresponding to ground-state structures at unseen compositions. The results demonstrate the NNP's ability to accurately reproduce energies under elastic lattice strains beyond the training scope, as well as make precise predictions of vibrational energies and forces for snapshots from MD simulations. Our investigation revealed the NNP's generalization accuracy trends when extrapolating to ground-state structures at unseen compositions. Notably, we observe a gradual degradation in performance as we move beyond the extensively trained composition range of the NNP model, especially when considering strained and vibrationally perturbed structural snapshots of ground states at far compositions.

%The high flexibility of NNP is the reason behind its better performance of NNP over CE can be due to two main reasons, the main reason is the much higher flexibility of NNP that allows large-scale adaptability to different energy basins in the PES. The other reason is the incorporation of geometric features, which makes it an off-lattice model capable of adapting to local geometric relaxations. which leads to better local approximation of the PES inside each energy basin. The former benefit was already observed in the ranking accuracy, while the latter was sensed when the NNP was used to extrapolate to the elastic and vibrational properties of the ground-state structures at unseen compositions. 

Additionally, we demonstrated an illustrative application of the NNP model for SA optimizations. Utilizing a multiple-cycle SA algorithm, we identified the low-energy ground-state and metastable structures that are anticipated to arise from experimental synthesis of Cr$_{(1-x)}$S phases in the composition range ($0.25 \leq x \leq 0.5$). Within this range, the NNP model accurately predicted the three known Cr-S ground-state phases listed in the Materials Project database \cite{jain2013commentary}: Cr$_{0.75}$S (mp-964), Cr$_{0.667}$S (mp-555569), and Cr$_{0.375}$S (mp-1181961). Moreover, the SA optimizations provided valuable insights into the preferred structures of Cr-S materials in this composition range. We observed a preference for structures that feature alternating CrS$_2$ and Cr-deficient layers, where the surplus Cr atoms beyond the ratio of Cr/S = 1/2 act as intercalants within the vdW gaps between CrS$_2$ layers. The presence of these intercalant Cr atoms in the vdW gaps significantly contributes to the stabilization of CrS$_2$ layers, even at relatively low percentages, down to around $0.03$. This becomes evident in the case of CrS$_2$ phase, where $50\;\%$ of the Cr atoms display a preferential migration to the vdW gaps to avoid the higher formation energy associated with the vdW CrS$_2$ phase.  

The NNP was further employed to investigate the influence of lateral compressive strain on CrS$_2$ slabs, resulting in a transition from non-vdW to vdW CrS$_2$. The analysis demonstrated the tendency of the slabs to release the strain-induced pressure through axial Poisson's expansion, giving rise to an alternating stress pattern within the slab layers. This alternating stress prompts the migration of Cr atoms from layers of high stress to those of low stress, leading to the formation of nearly empty vdW gaps between the CrS$_2$ layers. This behavior bears resemblance to successful endeavors in the CVD growth of vdW CrS$_2$ slabs, where a lattice mismatch of approximately $7\;\%$ arises due to the smaller lattice constant of the used f-mica substrate \cite{xiao2022van}. We conjecture that the smaller lattice parameter of the substrate is the reason behind the epitaxial growth of the vdW CrS$_2$ phase, rather than a non-vdW counterpart, wherein the anticipated compressive stresses due to lattice mismatch are released through vertical expansion of layers until stable vdW gaps are formed between the CrS$_2$ layers.

In light of the remarkable capability of the NNP model to provide energy and force predictions nearly as accurate as DFT, even for diverse and unobserved chemical compositions, atomic configurations, lattice strains, and atomic displacements, it holds great promise for steering off-lattice large-scale simulations of nonstoichiometric systems. As approximately $80\;\%$ of our dataset originates from previously idle structures in the CE training dataset, our methodology presents a cost-effective approach for modeling nonstoichiometric systems using NNPs. This approach holds practical implications in predicting phase diagrams, optimizing crystal structures, and simulating vacancy diffusion dynamics – a critical process in the synthesis of 2D materials using both top-down and bottom-up methods.  The kinetics of this diffusion process are intricately molded by the immediate atomic surroundings, which can undergo substantial stoichiometric variations over both spatial and temporal dimensions during materials synthesis. As a consequence, we advocate for the integration of NNPs or analogous MLPs for future computational studies on chemical exfoliation and epitaxial CVD growth of 2D materials. 

This work delved into the intricate chemical and geometrical degrees of freedom inherent in nonstoichiometric systems, with consideration only for the ground magnetic state obtained through spin-polarized DFT calculations. Prior studies have pioneered the combination of structural and spin degrees of freedom, enabling a comprehensive modeling that spans diverse magnetic configurations \cite{eckhoff2021high, novikov2022magnetic}. However, these models are currently limited to individual chemical configurations. Consequently, an avenue of considerable promise emerges for forthcoming studies to adeptly integrate magnetic spin degrees of freedom with the underlying chemical and geometrical attributes of nonstoichiometric systems. Such a holistic amalgamation holds the potential to significantly enhance the precision of modeling the synthesis processes of 2D magnetic materials, thereby expediting experimental mass production endeavors in this field. 

\section{Conclusions}

In conclusion, we have introduced a versatile framework that employs NNPs for modeling the nonstoichiometric characteristics inherent in the chemical exfoliation of non-vdW materials. We have convincingly demonstrated the superior performance of the NNP compared to the conventional CE model in terms of ability to bridge DFT accuracy to unseen crystal structures and compositions. By integrating the NNP into simulated annealing, we predicted the low-energy ground-state and metastable structures for diverse Cr$_{(1-x)}$S compositions. A key revelation was a structural shift at Cr$_{0.5}$S, with half the Cr atoms migrating to vdW gaps, underscoring the non-vdW nature of CrS$_2$ and highlighting the role of excess Cr beyond Cr/S = $1/2$ in vdW gap stabilization. Furthermore, we employed the NNP in an extensive vacancy diffusion MC simulation, illustrating the influential role of lateral compressive strain in promoting vdW gap formation in 2D non-vdW CrS$_2$ slabs. This finding suggests the potential of controlled strain engineering to enhance the efficiency of exfoliating ultrathin nanosheets from non-vdW crystals. In essence, our work establishes NNPs and analogous MLPs as invaluable tools for guiding computational simulations of chemical exfoliation of 2D non-vdW materials, with a specific emphasis on the intrinsic nonstoichiometric aspects of these processes.

\section{Supporting Information}
%Supplementary information is available for this paper.
To enhance reproducibility, the NNP training dataset, DFT, CE, and NNP settings files, along with Python codes for SA, MC vacancy diffusion simulations, and Jupyter notebooks for result analysis, are available on GitHub (https://github.com/UMBC-STEAM-LAB/NNP-nonstoichiometric-chromium-sulfides). 

\section{AUTHOR CONTRIBUTIONS}
AI trained and validated the NNP workflow, conducted DFT calculations for sampled structures, coded and executed all SA and MC simulations, and drafted the manuscript. DW conducted the DFT simulations of initial SQS cells sampled by CE and revised the manuscript. CA guided the project, oversaw all the work, and edited the manuscript.

\section{Acknowledgements}
The authors acknowledge the fund from the National Science Foundation (NSF) under grant number NSF DMR$-2213398$. 

\section{Notes}
The authors declare no competing financial interests. \\
%Please note that the use of the commercial software VASP does not imply recommendation by the National Institute of Standards and Technology.
Certain equipment, instruments, software, or materials are identified in this paper in order to specify the experimental procedure adequately. Such identification is not intended to imply recommendation or endorsement of any product or service by NIST, nor is it intended to imply that the materials or equipment identified are necessarily the best available for the purpose.

\section{REFERENCES}
\bibliography{main}% Produces the bibliography via BibTeX.

\end{document}

\section{Supplementary figures}

\maketitle

\begin{figure*}
\centering
\includegraphics[scale=0.9]{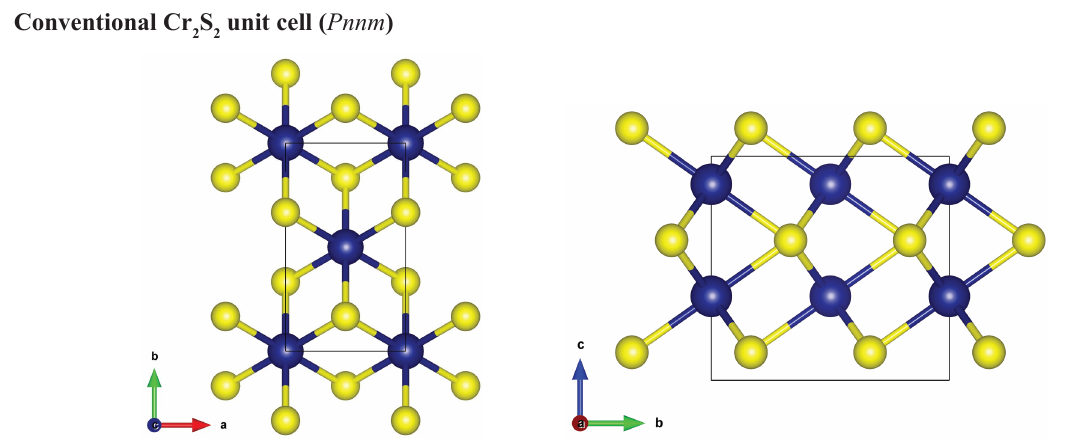}
\caption{\textbf{Conventional unit cell of the stoichiometric Cr-S phase (Cr$_2$S$_2$).} The DFT-optimized lattice parameters are $a = 3.486$ \AA{}, $b = 3.486 \times \sqrt{3}$ \AA{}, and $c = 5.750$ \AA{}.}
\label{fig:s1}
\end{figure*}

\begin{figure*}
\centering
\includegraphics[scale=0.2]{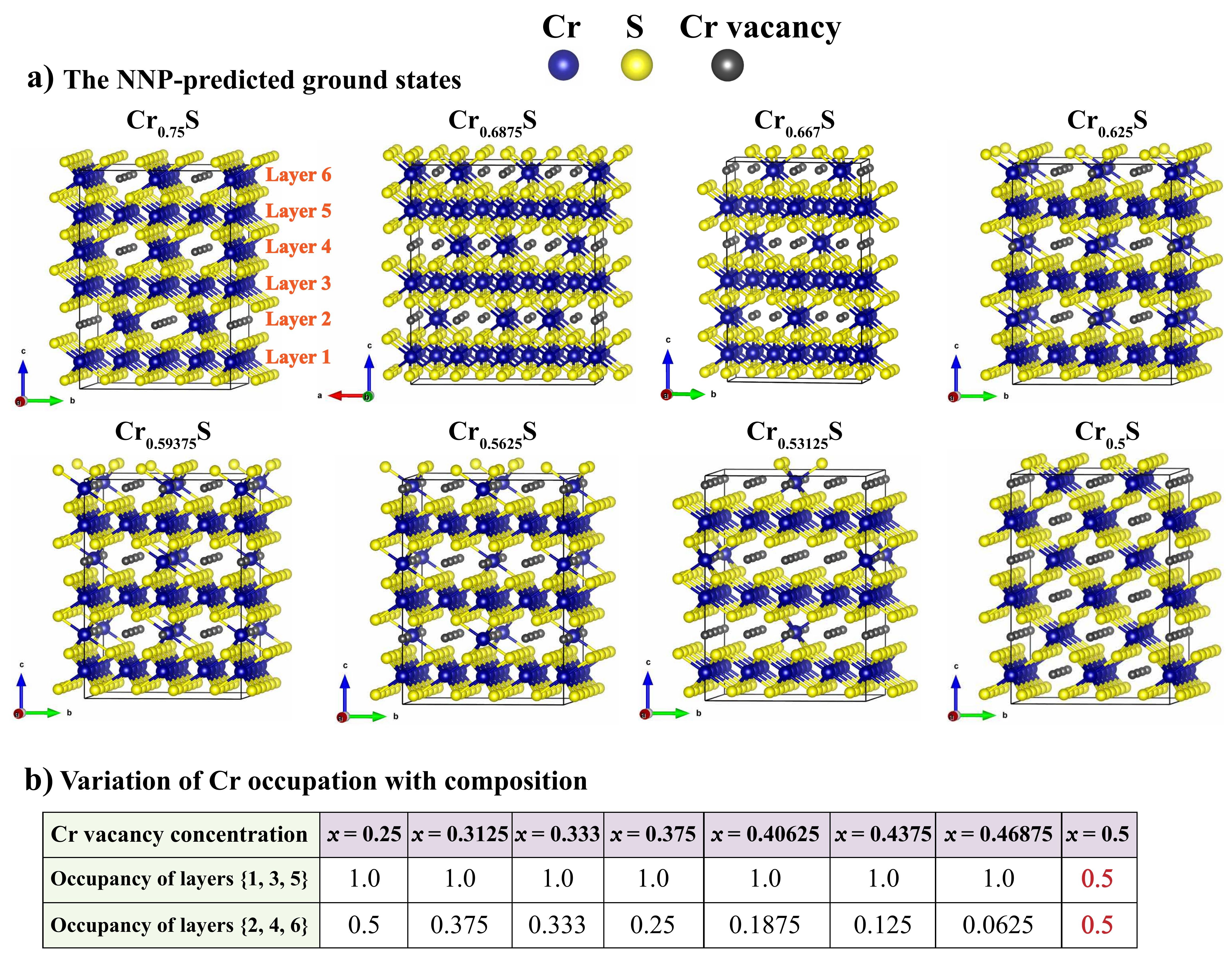}
\caption{\textbf{Simulated annealing prediction of Cr$_{(1-x)}$S ground-state structures.} a) illustrates the predicted ground states at $8$ compositions in the range $(0.25 \leq x \leq 0.5)$. b) depicts the occupation ratio of Cr sites with Cr atoms throughout the $6$ layers in the employed supercell, as a function of the Cr vacancy concentration.}
\label{fig:s2}
\end{figure*}

\begin{figure*}
\centering
\includegraphics[scale=0.2]{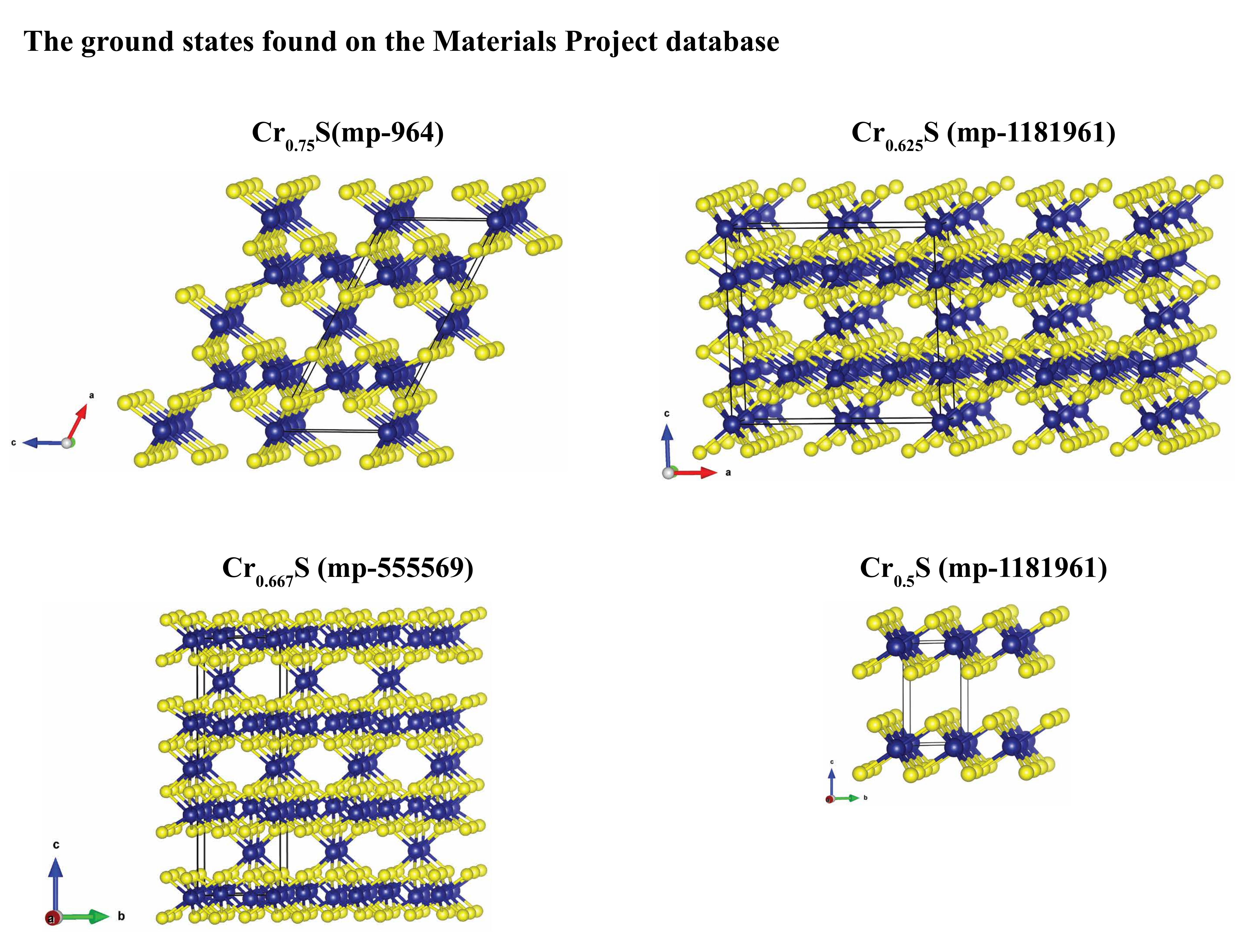}
\caption{The ground-state structures of Cr$_{(1-x)}$S in the composition range $(0.25 \leq x \leq 0.5)$ available on the Materials Project database.}
\label{fig:s3}
\end{figure*}

Figure \ref{fig:s1} illustrates the conventional unit cell of orthorhombic ($Pnnm$) Cr$_2$S$_2$. In Fig. \ref{fig:s2}, the ground-state structures resulting from the simulated annealing (SA) workflow, guided by the neural network potential (NNP), are depicted. For the SA process, we utilize supercells with dimensions of $4 \times 2\sqrt{3} \times 3$ based on the conventional unit cell in Fig. \ref{fig:s1}. In Fig. \ref{fig:s3}, we present the ground-state structures available in the Materials Project database for Cr$_{(1-x)}$S within the composition range of $(0.25 \leq x \leq 0.5)$ \cite{jain2013commentary}. We note that in order to accurately reproduce the known ground-state structure of Cr$_{0.667}$S (Cr$_2$S$_3$), an additional supercell configuration was considered. In this configuration, the zigzag orientation of the \textbf{a}-lattice vector in Fig. \ref{fig:s1} was rotated by $30^\circ$ to align with an armchair direction, forming a $60^\circ$ angle with the \textbf{b}-lattice vector. This adjustment was necessary as the initially employed conventional unit cell does not have the capability to generate the recognized Cr$_2$S$_3$ ground-state phase due to inherent symmetry constraints.

\begin{figure*}
\centering
\includegraphics[scale=0.7]{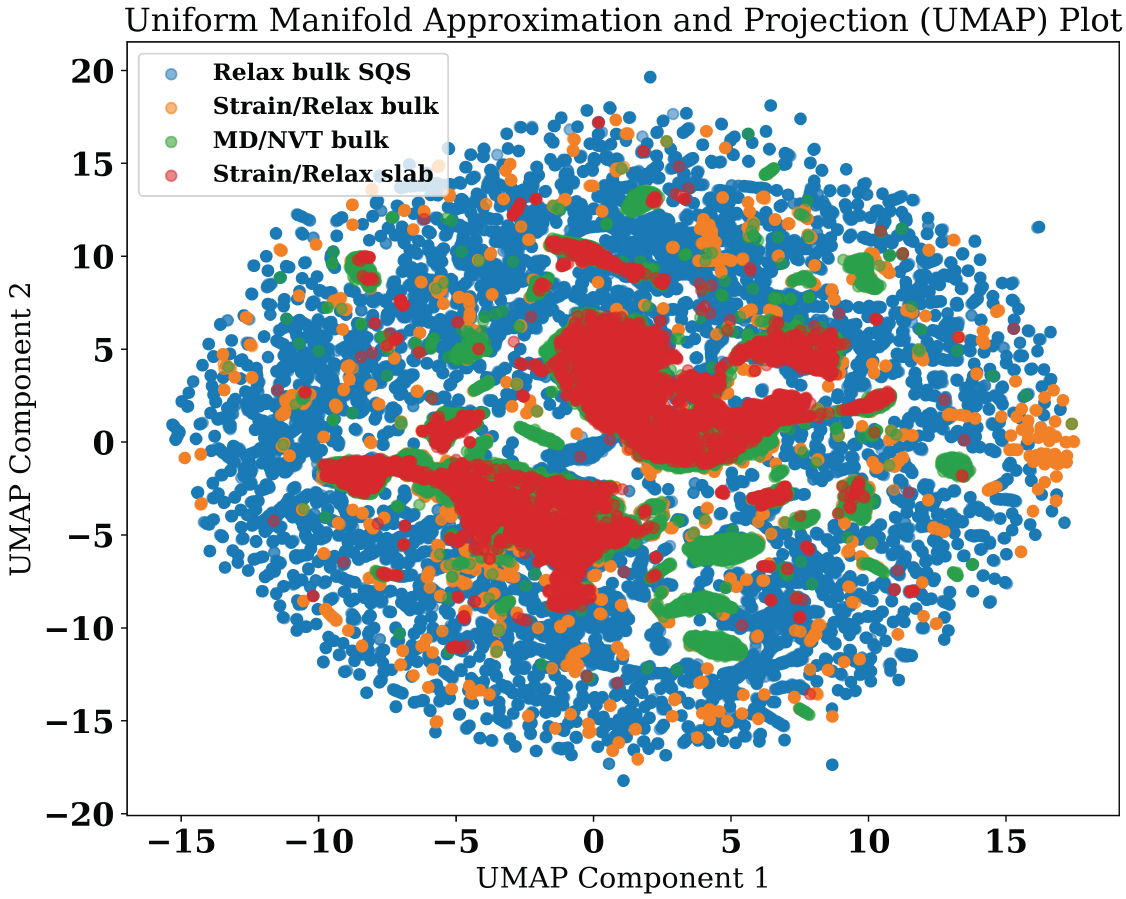}
\caption{Reduced-dimensionality representation of the features (symmetry functions) of all the atomic environments in the DFT (train/validation) dataset, obtained using the Uniform Manifold Approximation and Projection (UMAP) method.}
\label{fig:s4}
\end{figure*}

\begin{figure*}
\centering
\includegraphics[scale=0.7]{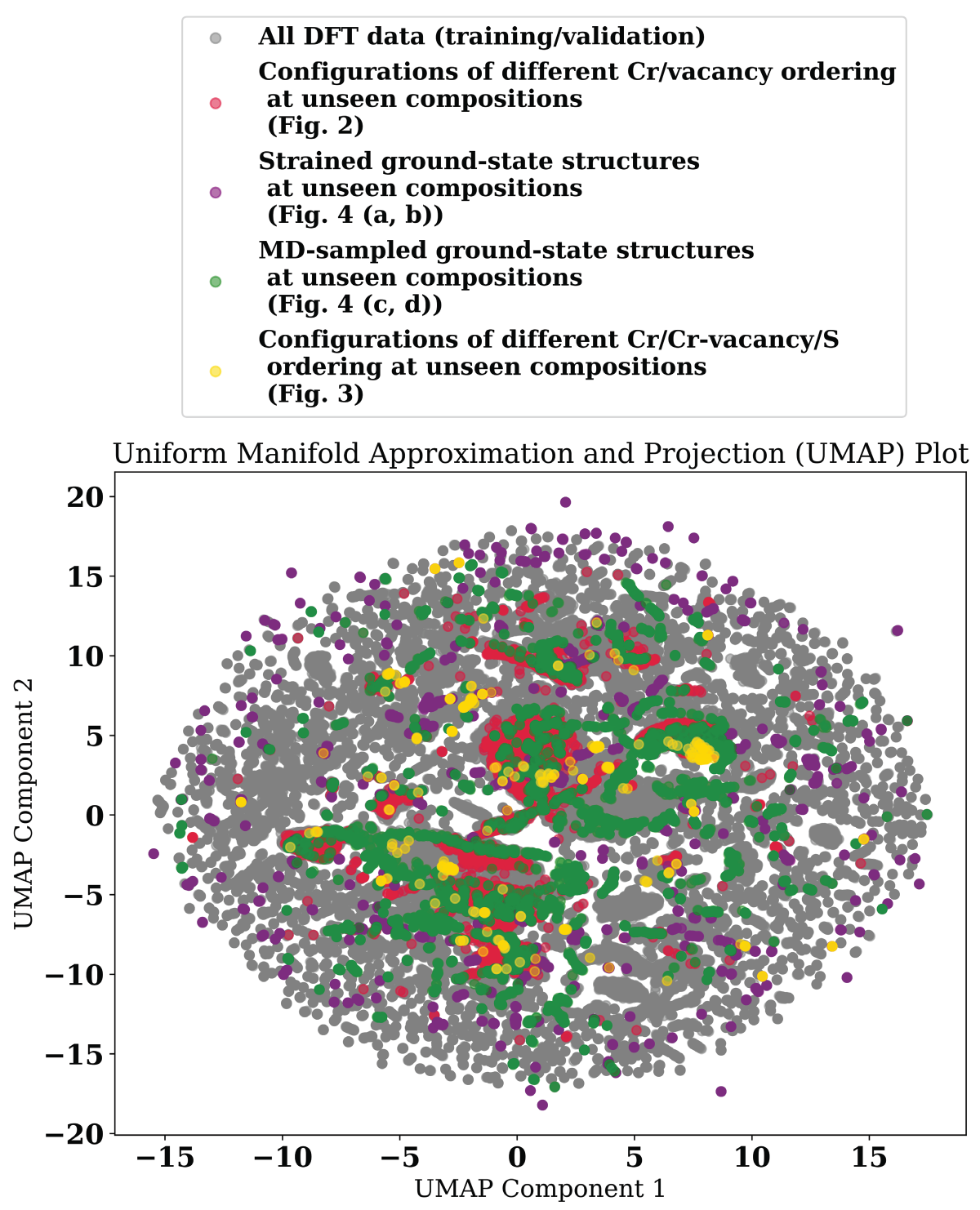}
\caption{Reduced-dimensionality representation of the features (symmetry functions) of the atomic environments of the unseen structures illustrated in Figures $2$, $3$, and $4$, used for evaluating the NNP generalizability. The DFT (train/validation) dataset representation is shown in the background for comparison.}
\label{fig:s5}
\end{figure*}

Figures \ref{fig:s4} and \ref{fig:s5} present a reduced-dimensionality representation of the feature space defined by the symmetry functions of the various atomic environments employed in this work, utilizing the Uniform Manifold Approximation and Projection (UMAP) method \cite{mcinnes2018umap}. The representation encompasses the distribution of the atomic environments within the DFT train/validation dataset, as well as the structures sampled to evaluate the generalization of the cluster expansion (CE) and the NNP models, discussed in (Results – Subsection A).

\begin{figure*}
\centering
\includegraphics[scale=0.45]{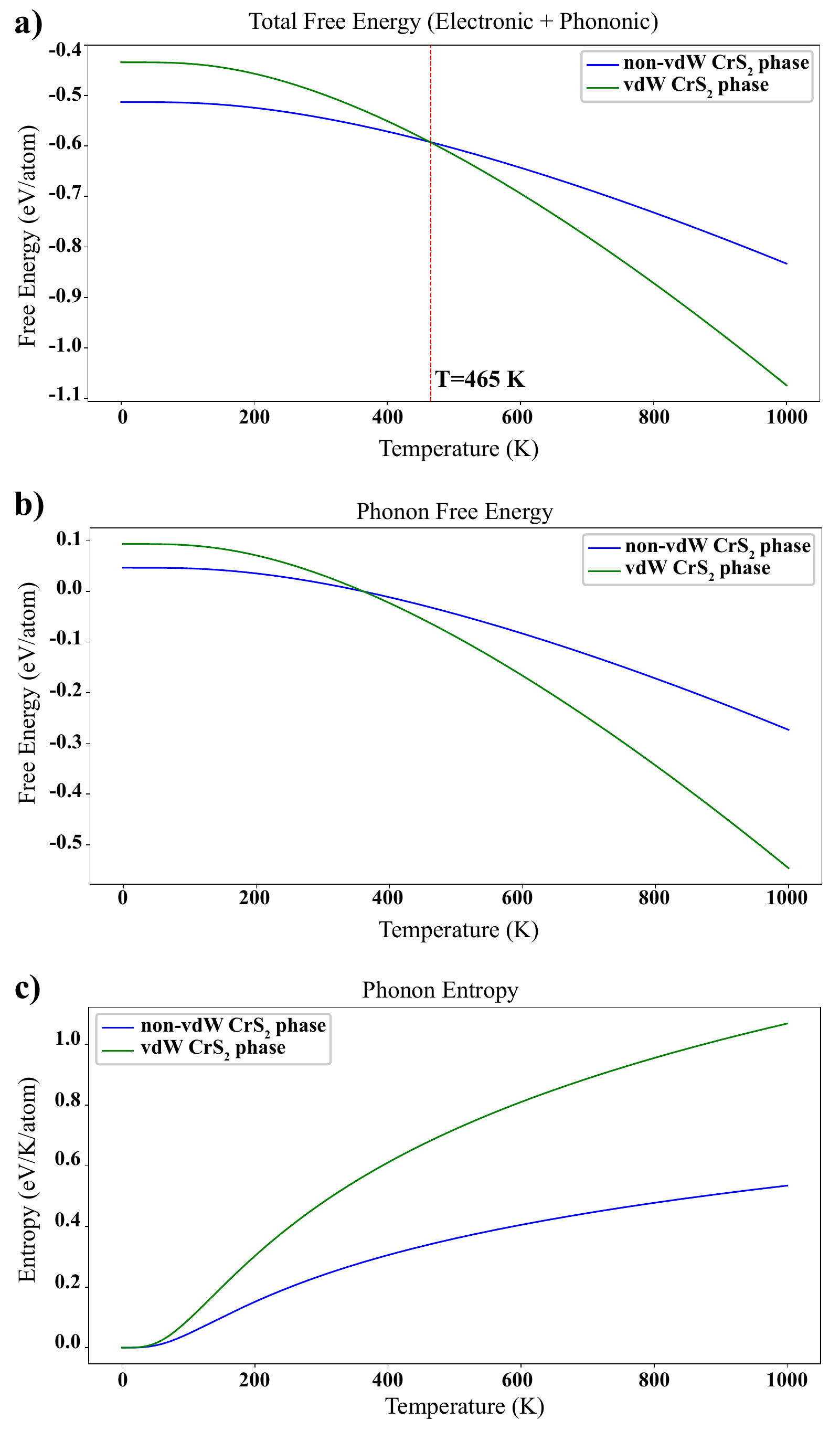}
\caption{\textbf{Stability of non-vdW vs vdW CrS$_2$.} a) Total free energy, b) vibrational (phononic) free energy, and c) phononic entropy.}
\label{fig:s6}
\end{figure*}

Figure \ref{fig:s6} illustrates the temperature-dependent relative stability between the van der Waals (vdW) and non-vdW CrS$_2$ phases discussed in (Results – Subsection B). The relative stability is indicated by the free energy $(F = E - TS)$, composed of the DFT electronic part at (T = $0$\;K) and the temperature-dependent vibrational free energy associated with phonons. The phonon contribution is calculated within the the harmonic approximation via the PHONOPY package \cite{phonopy-phono3py-JPCM, phonopy-phono3py-JPSJ}..

\clearpage
\section{Supplementary tables}

\begin{table}
\raggedright
\caption{\textbf{Details of the dataset:} N$_{SQS\;cells}$ refers to the number of bulk SQS cells sampled during CE fitting, while $N_{SQS\;strucs}$ denotes the number of structures sampled from SQS relaxation trajectories. $N_{GS\;bulk}$ and $N_{GS\;slab}$ signify the number of bulk and slab structures, respectively, generated from the identified ground states using strain/relaxation or strain/MD methods.}

\begin{tabular}{l c c c c}
    \toprule
    \multicolumn{1}{l}{Composition}  &
    \multicolumn{1}{c}{N$_{SQS\;cells}$} &
    \multicolumn{1}{c}{N$_{SQS\;strucs}$} &
    \multicolumn{1}{c}{N$_{{GS\;bulk}}$} &
    \multicolumn{1}{c}{N$_{{GS\;slab}}$} \\
    \midrule 
Cr$_{1.0}$S  & 1 & 6 & 10 & -- \\    
Cr$_{0.9}$S  & 2 & 30 & 77 & -- \\
Cr$_{0.875}$S & 11 & 173 & -- & -- \\
Cr$_{0.833}$S & 5 & 54 & 65 & -- \\
Cr$_{0.75}$S  & 12 & 185 & -- & -- \\
Cr$_{0.7}$S   & 9 & 255 & 499 & 186 \\
Cr$_{0.667}$S & 16 & 269 & 83 & 92 \\
Cr$_{0.625}$S & 26 & 654 & 515 & 139 \\
Cr$_{0.6}$S & 8 & 177 & -- & -- \\
Cr$_{0.5}$S & 48 & 1517 & 267 & -- \\
Cr$_{0.4}$S & 3 & 143 & -- & -- \\
Cr$_{0.375}$S  & 26 & 1044 & 74 & -- \\
Cr$_{0.333}$S & 16 & 494 & -- & -- \\
Cr$_{0.3}$S & 1 & 53 & -- & -- \\
Cr$_{0.25}$S & 24 & 1151 & 72 & -- \\
Cr$_{0.2}$S & 6 & 441 & -- & -- \\
Cr$_{0.167}$S & 6 & 281 & -- & -- \\
Cr$_{0.125}$S & 12 & 626 & -- & -- \\
Cr$_{0.1}$S & 5 & 277 & -- & -- \\
Cr$_{0.0714}$S & 4 & 539 & 80 & -- \\
Cr$_{0.0625}$S & 1 & 37 & -- & -- \\
Cr$_{0.0}$S & 1 & 18 & 10 & -- \\
    \bottomrule
\label{tab:s1}  
\end{tabular}

\end{table}

\clearpage

\begin{table}
\raggedright
\caption{\textbf{Details of CE fitting:} N$_{clusters}$ denotes the number of non-empty clusters employed in the CE basis, whereas N$_{atoms/cluster}$ lists the sizes of the considered clusters. E$_{train}$ and E$_{pred}$ respectively indicate the training RMSE and the prediction RMSE (obtained through N-fold CV) of the CE model. The energy errors are provided in meV/site, where "site" refers to Cr/vacancy active sites. The CPU time for CE calculation, on a single CPU core of Intel(R) Xeon(R) Gold 6230 @ 2.10GHz, is measured for the SQS cells in the dataset.}
  
\begin{tabular}{l c}
    \toprule
    \multicolumn{1}{l}{CE parameters}  &
    \multicolumn{1}{c}{} \\
    \midrule
        N$_{clusters}$ \;\;\;\;\;\;\;\;\;\;\;\;\;\;\;\;\;\;\;\;\;\;\;\;\;\;\;\;\;\;\;\;\;\;\;\;\;\;\;\;\;\;\;\;\;\;\;\;\;\;\;\;\;& 25 \\
        N$_{atoms/cluster}$ & 1, 2, 3 and 4 \\
        %N$_{sym\;op}$ & 24 \\
        E$_{train}$ RMSE (meV/site) & 80.3 \\
        E$_{CV}$ RMSE (meV/site) & 93.8 \\
        CPU time (ms/step/atom) & 3.31 ± 0.81 \\
    \bottomrule
\label{tab:s2}
\end{tabular}
\\

\end{table}

\begin{table}
\raggedright
\caption{\textbf{Details of the NNP fitting:} The NNP architecture consists of an input layer with $120$ descriptors for each structure, along with $2$ hidden layers, each comprising $24$ neurons. N$_{connections}$ represents the total count of weights and biases in the model. The training and prediction RMSE values are provided for both energies and forces, utilizing 80/20 train/test dataset split. The CPU time for NNP calculation, on a single CPU core of Intel(R) Xeon(R) Gold 6230 @ 2.10GHz, is measured for the SQS cells in the dataset.}
  
\begin{tabular}{l c}
    \toprule
    \multicolumn{1}{l}{NNP parameters}  &
    \multicolumn{1}{c}{} \\
    \midrule
        Architecture \;\;\;\;\;\;\;\;\;\;\;\;\;\;\;\;\;\;\;\;\;\;\;\;\;\;\;\;\;\;\;\;\;\;\;\;\;\;& $120-24-24-1$ \\
        N$_{connections}$ & 3529 \\
        R$_{cutoff}$ (\AA{}) & {2-body: 6.5, 3-body: 5.8} \\
        E$_{pred}$ RMSE (meV/atom) & 7.5 \\
        E$_{train}$ RMSE (meV/atom) & 8.7 \\
        F$_{pred}$ RMSE (eV/\AA{}) & 0.111 \\
        F$_{pred}$ RMSE (eV/\AA{}) & 0.124
        \\
        CPU time (ms/step/atom) & 3.95 ± 0.74 \\
    \bottomrule
\label{tab:s3}
\end{tabular}
\\

\end{table}

\bibliography{main}